\documentclass[sigconf]{acmart}
\renewcommand\footnotetextcopyrightpermission[1]{} 

\setcopyright{none}

\usepackage{graphicx}
\usepackage{enumerate}

\usepackage{amssymb}
\usepackage{amsmath}
\usepackage{bbm}
\usepackage{xfrac}
\usepackage{float}
\usepackage{verbatim}
\usepackage{xspace,afterpage}
\usepackage{ragged2e}
\usepackage{tablefootnote}
\usepackage{paralist}
\usepackage{tabularx}
\usepackage{subcaption}

\usepackage{dsfont}

\usepackage[ruled,vlined]{algorithm2e}
\usepackage{amsthm} 

\usepackage{soul}
            
\newtheorem{theo}{Theorem}

\newtheorem{cor}{Corollary}

\newtheorem{lem}{Lemma}

\AtBeginDocument{%
  \providecommand\BibTeX{{%
    \normalfont B\kern-0.5em{\scshape i\kern-0.25em b}\kern-0.8em\TeX}}}

\setcopyright{none}
\copyrightyear{none}
\acmYear{none}
\acmDOI{none}

\acmSubmissionID{\#34}

\begin{document}
\title{
LOAM: Low-latency Communication, Caching, and Computation Placement in Data-Intensive Computing Networks}

\author{Jinkun Zhang and Edmund Yeh}
\affiliation{%
  \institution{Northeastern University, United States}
  \country{}}
\email{jinkunzhang, eyeh @ece.neu.edu}



\begin{abstract}
Deploying data- and computation-intensive applications such as large-scale AI into heterogeneous dispersed computing networks can significantly enhance application performance by mitigating bottlenecks caused by limited network resources, including bandwidth, storage, and computing power. 
However, current resource allocation methods in dispersed computing do not provide a comprehensive solution that considers arbitrary topology, elastic resource amount, reuse of computation results, and nonlinear congestion-dependent optimization objectives.
In this paper, we propose LOAM, a low-latency joint communication, caching, and computation placement framework with a rigorous analytical foundation that incorporates the above aspects.
We tackle the NP-hard aggregated cost minimization problem with two methods: an offline method with a $1/2$ approximation and an online adaptive method with a bounded gap from the optimum. 
Through extensive simulation, the proposed framework outperforms multiple baselines in both synthesis and real-world network scenarios.
\end{abstract}





\maketitle
\section{Introduction}

Over the past decade, centralized clouds have been pivotal in IT service delivery. 
They are favored for their cost-effectiveness and ability to improve energy efficiency and computation speed for devices with limited resources \cite{satyanarayanan2017emergence}. 
However, the rise of Internet of Things (IoT) devices and the demand for services requiring ultra-low latency (e.g., online VR/AR gaming \cite{mangiante2017vr, wang2020user}, distributed learning on edge networks \cite{chen2021distributed}) have highlighted the limitations of centralized clouds, prompting a shift towards dispersed computing paradigms like fog and edge computing.
These paradigms distribute network resources, including bandwidth, storage, computing power, etc., closer to users, potentially surpassing centralized architectures in meeting the needs of delay-sensitive applications.
Dispersed computing platforms have gradually outperformed centralized cloud architectures in terms of request delay, scalability, error resilience, and AI/ML adaptation \cite{sriram2022edge}.
The move towards low-latency dispersed computing comes with its own set of challenges, particularly in managing network resources to handle data-intensive and computation-intensive (e.g., VR rendering \cite{mehrabi2021multi}) applications that demand large storage space and computing power.

Dispersed computing protocols address network resource allocation control with different granularity.
For example, in fog computing (FC), networking, storage, and computing resources are distributed across hierarchical levels from the backbone to the edge \cite{hu2017survey}.
In mobile edge computing (MEC), resources are distributed
throughout the mobile edge close to users.
Task offloading is considered at the application layer \cite{islam2021survey}, and caching at the network layer \cite{zeydan2016big}. 
By jointly considering forwarding (on what paths are requests routed), caching (what to put in network storage), and computation placement (where to offload computation jobs) in a cross-layer and mathematically rigorous manner, this paper is dedicated to further pushing dispersed computing performance to its limits.

On the one hand, current dispersed computing resource allocation algorithms focus on various architectures (e.g., Collaborative Edge Computing \cite{ning2018green}, mesh networks \cite{apostolaras2016mechanism}, Internet of Things \cite{huang2021task}, geo-distributed learning \cite{hsieh2017gaia}) and performance metrics (e.g., service delay \cite{liu2016delay}, network throughput \cite{kamran2021deco}, fairness \cite{zhao2021fairness}).
Nevertheless, most of the previous algorithms are optimized for specific network topologies (e.g., two-hop networks) and fixed resource amounts (e.g., total computing power or storage size). 
Whereas in future collaborative networks, dispersed computing systems are expected to adapt to arbitrary topologies and elastic network resources.
For example, next-gen large-scale AI could emerge with device-edge-cloud fusion \cite{li2022design}, and network operators could make trade-offs between resource efficiency and performance to maximize revenue\cite{cooper2021accuracy}. 
Moreover, with the explosive growth of smart devices and online applications, congestion mitigation has become a crucial design objective for fog/edge networks. 
Frameworks that optimize for linear costs with link/CPU capacity constraints \cite{liu2019joint} or network throughput \cite{kamran2021deco} could suffer from suboptimality in terms of latency since the queueing effect caused by congestion is not reflected in the model. 
In contrast, this paper directly considers non-linear congestion-dependent queueing delay on links/CPUs.

On the other hand, previous works have efficiently optimized for forwarding \cite{zhou2021energy}, task offloading \cite{WiOpt22} and caching \cite{xia2020online} separately.
Numerous joint optimization frameworks have also been proposed recently, including joint offloading and bandwidth \cite{huang2019deep}, joint caching and wireless power \cite{malak2023joint}, joint caching and offloading \cite{bi2020joint}, joint forwarding, offloading, and caching \cite{kamran2021deco}.
With the increase of data- and computation-intensive applications, however, existing methods can suffer from suboptimal interaction between computing and storage utilization.
For instance, a throughput-optimal task offloading and caching algorithm proposed by \cite{kamran2021deco} intelligently allocates computation and data storage.
Nevertheless, when requests for the same computation on the same data are frequently generated concurrently (e.g., when multiple VR users make rendering requests for the same Point of View), directly caching computational results can significantly reduce request latency compared to purely moving computation and data closer to each other.
To this end, \emph{computation reuse} (i.e., reuse of computational results) has been proposed and recognized as an important feature \cite{barrios2023service, al2022promise, cui2017energy,mastorakis2020icedge}. However, it is often considered separate from other network resources and thus does not provide a unified analytical performance guarantee.

Combining the above, to our knowledge, no previous works archived latency-minimal resource allocation for cache-enabled dispersed computing with arbitrary topology, elastic resource amounts, computation reuse, and nonlinear congestion-dependent objectives.
In this paper, we fill this gap by proposing a conceptually novel and analytically rigorous framework named \emph{LOAM} (low-Latency cOmmunication, cAching, and coMputation) that efficiently manages the forwarding, caching and offloading strategies for cache-enabled computing networks in a distributed manner.
It achieves constant factor guarantees for minimizing general nolinear costs in arbitrary heterogeneous networks, and incorporates elastic resource allocation and computation reuse.
The idea of LOAM can be applied in a number of cutting-edge data- and computation-intensive network paradigms, e.g., IoT-enabled healthcare \cite{selvaraj2020challenges}, Edge-AI \cite{wang2019edge}, scientific experimental networks \cite{wu2022n}, and in-network video processing \cite{sun2019mvideo}.

Specifically, LOAM considers an arbitrary multi-hop network, where nodes have heterogeneous capabilities for communication, computation, and storage. 
Nodes can issue \emph{computation requests} for performing computation tasks on network-cached data (e.g., inference using a trained model saved in the network).
Computation requests are offloaded to nodes with adequate computing power, and corresponding \emph{data requests} are issued to fetch required data back to the computing node for computation.
Since nodes are equipped with elastic-size caches, they can temporarily store data and/or computation results to boost task completion. 
In this network, nonlinear costs (e.g., queueing delay on links/CPUs, and cost for cache deployment) are incurred on the links and nodes due to transmission, computation, and caching. 
LOAM provides an integrated solution for aggregated cost minimization in a distributed manner, jointly over forwarding, offloading, and caching.  

LOAM tackles this NP-hard optimization problem with a solid analytical foundation. It provides an offline algorithm and an online adaptive algorithm, both with performance guarantees.
Specifically, the offline solution archives a $1/2$ approximation to an equivalent problem by exploiting its ``submodular $+$ concave'' structure, however, requiring prior knowledge of network status and request pattern; 
the online solution guarantees a bounded gap from the optimum using the problem's ``convex $+$ geodesic-convex'' structure, and can adapt to changes in request pattern and network status without prior knowledge.
To our knowledge, LOAM is the first to provide performance guarantees to this problem.
LOAM demonstrates its advantage over multiple baselines in synthesis and real-world scenarios through extensive packet-level simulations.

The key contributions of this paper are summarized as follows:
\begin{itemize}
    \item We are the first to formulate heterogeneous cache-enabled computing networks with arbitrary topology, elastic resources, computation reuse, and nonlinear costs.
    \item We provide the first performance guarantee for cache-enabled computing networks with arbitrary topology and nonlinear costs: we provide an offline solution to the NP-hard problem with $\frac{1}{2}$ approximation, and devise an online and adaptive solution to the problem by a modified KKT condition that guarantees a bounded gap from the optimum.
    \item We provide a public-available packet-level network simulator. Extensive simulation shows that proposed solutions significantly outperform baselines in multiple scenarios.
\end{itemize}



\section{Network model}
\label{sec:model}

\subsection{Cache-enabled computing network}
\label{subsec:computing network}
Consider a network modeled by a directed graph $\mathcal{G} = (\mathcal{V},\mathcal{E})$, where $\mathcal{E}$ is the set of directed links, and $\mathcal{V}$ is the set of nodes, each capable of performing computing tasks, caching data or computational result objects, and communicating with other nodes.
We assume $(j,i) \in \mathcal{E}$ if $(i,j) \in \mathcal{E}$.
For $i \in \mathcal{V}$, let $\mathcal{N}(i) = \left\{j \in \mathcal{V} \big| (i,j) \in \mathcal{E}\right\}$ denote the neighbors of node $i$.
We assume there is a \emph{computation catalog} $\mathcal{F}$ for the computations that can be performed in $\mathcal{G}$, and a \emph{data catalog} $\mathcal{C}$ for data objects in $\mathcal{G}$, where $\mathcal{F}$ and $\mathcal{C}$ are finite.\footnote{We briefly discuss the case of infinite data or computation catalog in Section \ref{subsec:GeneralAlgorithm}.}
Network $\mathcal{G}$ is \emph{task-driven}.
A task is denoted by a 3-tuple $(d,m,k)$, where $d \in \mathcal{V}$ is the task requester (i.e., the destination to which computation results will be delivered), $m \in \mathcal{F}$ is desired the computation and $k \in \mathcal{C}$ is the required data object.
\footnote{
We assume the ``computation'' specified by $m$ includes user-specified computational inputs. e.g., $k$ may refer to a VR 3D model, and $m$ may refer to rendering it with a specific Point-of-View (PoV). Different PoVs are represented by different $m$.}
To guarantee the feasibility of tasks, for each $k \in \mathcal{C}$, there is a non-empty set of \emph{designated servers} denoted by $\mathcal{S}_k \subseteq \mathcal{V}$ that keeps $k$ permanently.

For task $(d,m,k)$, we assume $d$ generates \emph{computation interests} (CI) packets for computation $m$ on data $k$ at rate $r_d(m,k)$ (packet/sec).
With the absence of data/result caching, CI packets are routed in $\mathcal{G}$ hop-by-hop. 
Nodes receiving a CI can decide whether to perform the computation locally or forward it to other nodes.
Once a node $i$ decides to perform the computation, it puts the CI on pending and generates a corresponding \emph{data interest} (DI) for $k$.
DI packets are also forwarded hop-by-hop until reaching designated servers.
Server responds to DI with a \emph{data response} (DR) packet that carries $k$ back to $i$. 
Then $i$ can resolve the pending CI by performing computation $m$ on $k$. 
After computation, it sends a \emph{computation response} (CR) packet that delivers the computational result to requester $d$.
We assume the CR and DR packets are routed on the same paths as CI and DI but in the reverse direction, respectively.

\begin{figure}[htbp]
\centerline{\includegraphics[width=1\linewidth]{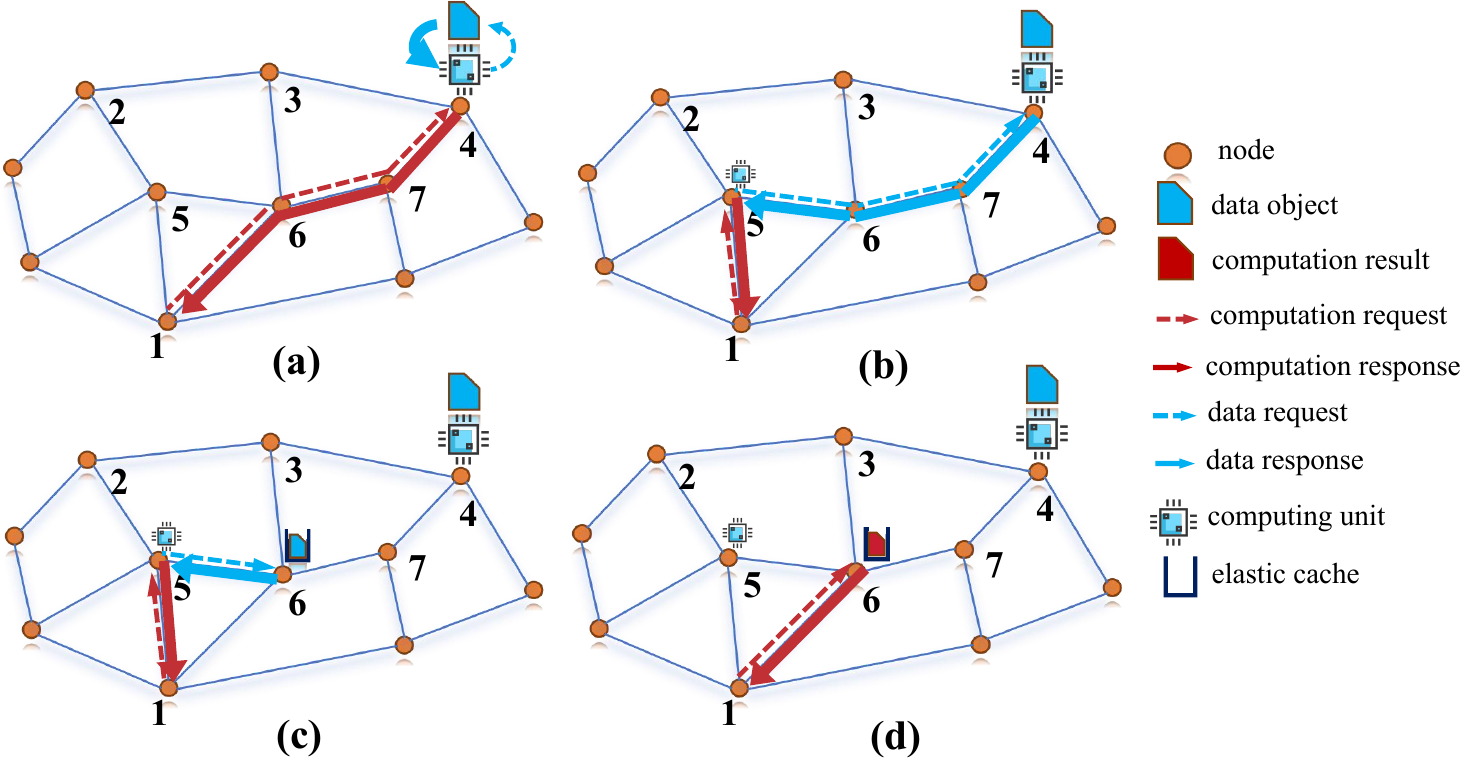}}
\caption{(a) Could computing, server stores all data and performs all computations. (b) Offloading computations, retrieving data from server \cite{WiOpt22}. (c) Offloading computations and caching data objects \cite{kamran2021deco}. (d) Reusing computation results.}
\label{fig_network}
\end{figure}

The above mechanism is boosted by utilizing in-network storage. 
We assume nodes in $\mathcal{V}$ are equipped with caches of elastic sizes.
The cache can temporarily store data objects and/or computational results.
Specifically, upon receiving a CI, a node first checks if the result is cached. If so, the corresponding CR is generated and sent right away without any computation performed.
Upon receiving a DI, the corresponding DR is generated and sent if the required data object is cached.
\footnote{The implementation and formatting details of CI, CR, DI and DR packets in practical systems are omitted for brevity. We refer the readers to works on Named-Data Networking \cite{zhang2014named} and Named-Function-Networking \cite{krol2017nfaas}.}
This paper is the first to minimize latency by reusing computational results in arbitrary data-centric computing networks. Fig.~\ref{fig_network} illustrates our network model.

\subsection{Node-based strategies}

We aim to optimize the time-averaged network performance via a node-based formulation. 
Let $x_i^\text{d}(k) \in \{0,1\}$ be the \emph{cache decision} of node $i$ for data $k$, i.e., $x_i^\text{d}(k) = 1$ if $i$ caches $k$ and $0$ if not, and let $x_{i}^\text{c}(m,k) \in \{0,1\}$ be the cache decision of node $i$ for the computational result of computation $m$ on data $k$.
We denote by $\boldsymbol{x} = [x_i^\text{d}(k), x_{i}^\text{c}(m,k)]_{i\in\mathcal{V},k \in \mathcal{C}, m\in\mathcal{F}}$ the \emph{global cache decision}.
We assume hop-by-hop multipath routing in $\mathcal{G}$.
If $x_{i}^\text{c}(m,k) = 0$, of the CI packets received by node $i$ for computation $m$ on data $k$, a fraction $\phi^{\text{c}}_{ij}(m,k) \in [0,1]$ is forwarded to neighbor node $j \in \mathcal{N}(i)$, and a fraction $\phi^{\text{c}}_{i0}(m,k) \in [0,1]$ is put on pending for local computation.
If $x^\text{d}_i(k) = 0$ and $i \not\in \mathcal{S}_k$, of the DI packets for data $k$ received by $i$, a fraction $\phi^{\text{d}}_{ij}(k) \in [0,1]$ is forwarded to neighbor $j$.\footnote{We assume $\phi^{\text{c}}_{ij}(m,k)=\phi^{\text{d}}_{ij}(k)=0$ if $(i,j) \not\in \mathcal{E}$, and $\phi^{\text{d}}_{ij}(k) = 0$ for all $j$ if $i \in \mathcal{S}_k$.}
We denote by 
$\boldsymbol{\phi} = [\phi^{\text{c}}_{ij}(m,k), \phi^{\text{d}}_{ij}(k)] _{i,j \in \mathcal{V}, k \in \mathcal{C}, m \in \mathcal{F}}$
the \emph{global forwarding strategy}.
The following holds for all $i \in \mathcal{V}$,
\begin{small}
\begin{equation}
\label{flow_conservation}
\begin{aligned}
    \sum\nolimits_{j \in \mathcal{V}\cup\{0\}} \phi^{\text{c}}_{ij}(m,k) + x^{\text{c}}_{i}(m,k) = 1, \quad \forall m \in \mathcal{F}, k \in \mathcal{C},
    \\ \sum\nolimits_{j \in \mathcal{V}} \phi^{\text{d}}_{ij}(k) + x^{\text{d}}_{i}(k) = \begin{cases}
        0, \quad \text{if } i \in \mathcal{S}_k
        \\ 1, \quad \text{if } i \not\in \mathcal{S}_k
    \end{cases}  \forall k \in \mathcal{C}.
\end{aligned}
\end{equation}
\end{small}
Conservation \eqref{flow_conservation} implies each CI (DI) packet is forwarded to one of the neighbor nodes if the corresponding result (data) is not stored at the present node. 
It guarantees that all computation requests are fulfilled (i.e., each CI will eventually lead to a corresponding CR).

\begin{figure}[htbp]
\centerline{\includegraphics[width=1\linewidth]{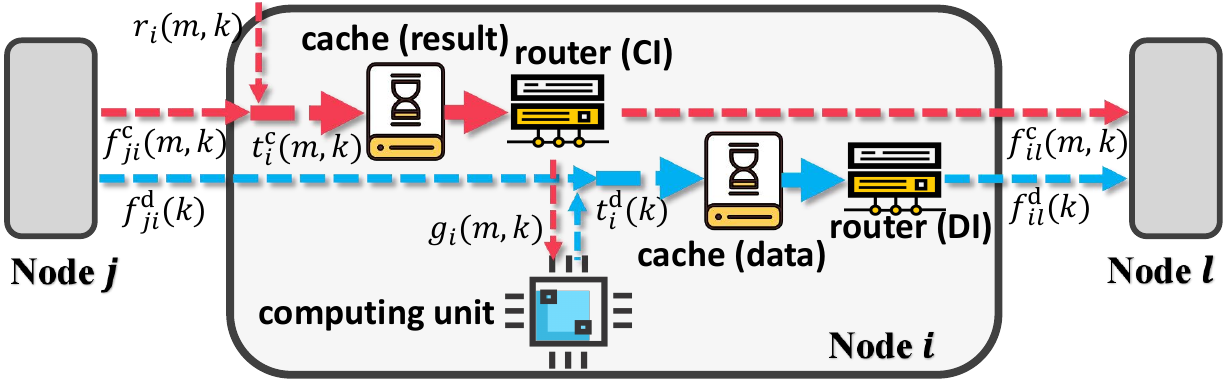}}
\caption{Flow level behavior of nodes $j \to i \to l$. 
We only mark the flows of CI and DI. Flows of CR/DR are on the same path as CI/DI, in the reversed direction.}
\label{fig_flow}
\end{figure}

Let $t^{c}_i(m,k)$ be node $i$'s time-averaged \emph{traffic} (packet/sec) for CI packets of computation $m$ on data $k$.
It includes CI packets both generated by $i$ and forwarded from neighbor nodes to $i$, namely,
\begin{subequations}
\label{traffic}
\begin{small}
\begin{equation}
    t^{c}_i(m,k) = r_i(m,k) + \sum\nolimits_{j \in \mathcal{N}(i)}  \phi^{\text{c}}_{ji}(m,k) t^{c}_j(m,k).
\end{equation}
\end{small}
We denote by $g_i(m,k)$ the average number of computation $m$ performed by node $i$ on data $k$, where $g_i(m,k) = t^{\text{c}}_i(m,k) \phi^{\text{c}}_{i0}(m,k)$.
Let $t^\text{d}_i(k)$ be $i$'s traffic of DI for data $k$. It includes DI both generated by $i$ (due to pending CI), and forwarded from other nodes, i.e.,
\begin{small}
\begin{equation}
    t^{d}_i(k) = \sum\nolimits_{m \in \mathcal{F}} g_i(m,k) + \sum\nolimits_{j \in \mathcal{N}(i)}  \phi^{\text{d}}_{ji}(k) t^{d}_j(k).
\end{equation}
\end{small}
\end{subequations}
Fig~\ref{fig_flow} gives a detailed illustration of our flow model.
We remark that our decision variables are only $\boldsymbol{x}$ and $\boldsymbol{\phi}$. For a given $\boldsymbol{\phi}$, the uniqueness of $t^{c}_i(m,k)$ and $t^{d}_i(k)$, i.e., the existence of a unique solution to \eqref{traffic}, is shown in \cite{gallager1977minimum} Theorem $1$.

\subsection{Problem formulation}
Nonlinear costs are incurred on the links due to packet queueing/transmission, and at the nodes due to computation and cache deployment.
Let $f_{ij}^\text{c}(m,k)$ be the rate (packet/sec) of CI for computation $m$ and data $k$ that travel through $(i,j)$. 
Let $f_{ij}^\text{d}(k)$ be the rate of DI for data $k$ on $(i,j)$. It holds that $f_{ij}^\text{c}(m,k) = t_i^{\text{c}}(m,k) \phi^{\text{c}}_{ij}(m,k)$, and $f_{ij}^\text{d}(k) = t_i^{\text{d}}(k) \phi^{\text{d}}_{ij}(k)$.
Since the size of interest packets (CI/DI) is typically negligible compared to that of response packets (CR/DR) carrying computational results and data objects, we only consider the link cost caused by CR/DR.
Let $L_k^{\text{d}}$ (bit) be the size of data $k \in \mathcal{C}$, and let $L_{mk}^{\text{c}}$ be result size of computation $m\in\mathcal{F}$ on $k$.
Since \eqref{flow_conservation} implies every CI/DI on $(i,j)$ must be replied with a corresponding CR/DR on $(j,i)$, the link flow rate (bit/sec) on $(i,j)$ is given by
\begin{small}
\begin{equation*}
    F_{ij} = \sum\nolimits_{m\in\mathcal{F}, k \in \mathcal{C}} L_{mk}^{\text{c}} f_{ji}^\text{c}(m,k) + \sum\nolimits_{k \in \mathcal{C}}L_k^{\text{d}}f_{ji}^\text{d}(k).
\end{equation*}
\end{small}
We denote by $D_{ij}(F_{ij})$ the link cost on $(i,j)$, and assume $D_{ij}(\cdot)$ is continuously differentiable, monotonically increasing and convex, with $D_{ij}(0) = 0$.
Such $D_{ij}(\cdot)$ is commonly adopted \cite{gallager1977minimum,xi2008node} as it subsumes a variety of existing cost functions, including linear transmission delay and hard link capacity constraints. 
It also incorporates congestion-dependent metrics. 
Let $\mu_{ij}$ be the service rate of an M/M/1 queue, then $D_{ij}(F_{ij}) = {F_{ij}}/\left({\mu_{ij}-F_{ij}}\right)$ gives the average number of packets waiting in the queue or being served \cite{bertsekas2021data}. 

To measure computation costs, let $W_{imk}$ be the workload for node $i$ to perform a single computation request of $m$ on $k$, the total computational workload at node $i$ is given by
\begin{small}
\begin{equation*}
    G_i = \sum\nolimits_{m \in \mathcal{F}, k \in \mathcal{C}} W_{imk} g_i(m,k).
\end{equation*}
\end{small}
The computation cost at $i$ is denoted by $C_i(G_i)$, where $C_i(\cdot)$ is also increasing, continuously differentiable and convex, with $C_i(0) = 0$.
Function $C_i(\cdot)$ can incorporate computation congestion (e.g., average number of packets waiting for available processor or being served at CPU).
By Little's Law, when both $D_{ij}(F_{ij})$ and $C_i(G_i)$ represent queue lengths, $\sum_{(i,j)\in\mathcal{E}}D_{ij}(F_{ij}) + \sum_{i \in \mathcal{V}}C_i(G_i)$ is proportional to the expected system latency of computation tasks.

Elastic storage spaces caching data objects or computational results also introduce costs. 
Let $X_i$ be the cache size at node $i$,
\begin{small}
\begin{equation*}
    X_i = \sum\nolimits_{m \in \mathcal{F}, k \in \mathcal{C}}x^{\text{c}}_i(m,k) + \sum\nolimits_{k \in \mathcal{C}}x^{\text{d}}_i(k).
\end{equation*}
\end{small}
We denote by $B_{i}(X_{i})$ the cache deployment cost at node $i$, where $B_{i}(\cdot)$ is also continuously differentiable, monotonically increasing and convex, with $B_{i}(0) = 0$.
It can represent the expense to buy/rent storage \cite{ye2021joint,dehghan2019utility}, utility measured by expenses and read/write speed \cite{chu2018joint}, or approximate hard cache capacity constraints \cite{ioannidis2016adaptive}.

We wish to jointly tune $\boldsymbol{x}$ and $\boldsymbol{\phi}$ to minimize the network aggregated cost for transmission, computation, and caching. 
To construct a continuous relaxation to the mixed-integer problem, suppose $x_i^\text{c}(m,k)$ and $x^\text{d}_{i}(k)$ are independent Bernoulli random variables. 
Let $\nu$ be the corresponding joint probability distribution over matrices in $\{0,1\}^{|\mathcal{V}|\times|\mathcal{F}|\times |\mathcal{C}|}$ and $\{0,1\}^{|\mathcal{V}|\times|\mathcal{C}|}$, and denote by $\boldsymbol{P}_{\nu}[\cdot]$, $\mathbb{E}_{\nu}[\cdot]$ the probability and expectation w.r.t. $\nu$.
Let $y^\text{c}_i(m,k) \in [0,1]$ be the probability that $i$ caches result for computation $m$ on $k$, i.e., $y^\text{c}_i(m,k) = \boldsymbol{P}_{\nu}[x^\text{c}_i(m,k) = 1] = \mathbb{E}_{\nu}[x^\text{c}_i(m,k)]$.
Let $y^\text{d}_i(k) = \boldsymbol{P}_{\nu}[x^\text{d}_i(k) = 1] = \mathbb{E}_{\nu}[x^\text{d}_i(k)]$ be the probability that node $i$ caches data $k$. 
Let $\boldsymbol{y} = [y_i^\text{d}(k), y_{i}^\text{c}(m,k)]_{i\in\mathcal{V},k \in \mathcal{C}, m\in\mathcal{F}}$ be the \emph{global caching strategy}. 
Take expectation w.r.t. $\nu$, conservation \eqref{flow_conservation} becomes
\begin{small}
\begin{equation}
\label{flow_conservation_relax}
\begin{aligned}
    \sum\nolimits_{j \in \{0\}\cup\mathcal{V}} \phi^{\text{c}}_{ij}(m,k) + y^{\text{c}}_{i}(m,k) = 1, \quad \forall m \in \mathcal{F}, k \in \mathcal{C},
    \\ \sum\nolimits_{j \in \mathcal{V}} \phi^{\text{d}}_{ij}(k) + y^{\text{d}}_{i}(k) = \begin{cases}
        0, \quad \text{if } i \in \mathcal{S}_k
        \\ 1, \quad \text{if } i \not\in \mathcal{S}_k
    \end{cases}  \forall k \in \mathcal{C}.
\end{aligned}
\end{equation}
\end{small}

Without ambiguity, we use $F_{ij}$ to denote $F_{ij}|_{\boldsymbol{y}} = \mathbb{E}_{\nu}[F_{ij}]$ in the rest of the paper, and let $Y_i = \sum_{k \in \mathcal{C}}y_i(k)$ denote the expected cache size.
The aggregated cost minimization problem is cast as
\begin{equation}
    \begin{aligned}
        \min_{\boldsymbol{y},\boldsymbol{\phi}} \, T(\boldsymbol{y},\boldsymbol{\phi})& = \sum_{(i,j) \in \mathcal{E}} D_{ij}(F_{ij}) + \sum_{i\in\mathcal{V}}B_{i}(Y_{i}) + \sum_{i\in\mathcal{V}}C_{i}(G_{i})
        \\\text{subject to} \quad 
        & \phi^{\text{c}}_{ij}(m,k) \in [0,1], \quad \phi^{\text{d}}_{ij}(k) \in [0,1], 
        \\ & y^{\text{c}}_{i}(m,k) \in[0,1], \quad y^{\text{d}}_{i}(k) \in [0,1],
        \\ & \text{ \eqref{flow_conservation_relax} holds.}
    \end{aligned}
    \label{Objective_relax}
\end{equation}

Note that we do not explicitly impose any constraints for link, computation, or cache capacities in \eqref{Objective_relax}, as any capacity limitation can be incorporated in the convex cost functions.
In this paper, we present two methods to tackle \eqref{Objective_relax}: an offline centralized method in Section \ref{sec: submodular} and an online distributed method in Section \ref{sec:general condition}. 
We summarize the major notations of this paper in Table~\ref{table:notation}.

\begin{table}[t]
\begin{tabular}{l | l }
\hline
$\mathcal{V}, \mathcal{E}$ & set of nodes and directed links\\
$\mathcal{F}, \mathcal{C}$ & set of available computations and data objects
\\
$\mathcal{S}_k$ & set of designated servers for data $k \in \mathcal{C}$\\
$r_i(m,k)$ & CI input rate for computation $m$ on data $k$ at node $i$ \\
$\mathcal{T}$ & set of all tasks \\
$t^\text{c}_i(m,k)$ & CI traffic for computation $m$ on data $k$ at node $i$\\
$t^\text{d}_i(k)$ & DI traffic for data $k$ at node $i$\\
$f^\text{c}_{ij}(m,k)$ & rate of CI packets for $m,k$ forwarded from $i$ to $j$\\
$f^\text{d}_{ij}(k)$ & rate of DI packets for $k$ forwarded from $i$ to $j$\\
$g_i(m,k)$ & rate of CI for $m,k$ for local computation at $i$\\
$\phi^\text{c}_{ij}(m,k)$ & fraction of CI traffic at $i$ that are forwarded to $j$\\
$\phi^\text{c}_{i0}(m,k)$ & fraction of CI traffic at $i$ for local computation\\
$\phi^\text{d}_{ij}(k)$ & fraction of DI traffic at $i$ forwarded to $j$\\
$y^\text{c}_i(m,k)$ & caching strategy of computation result for $m,k$ at $i$ \\
$y^\text{d}_i(k)$ & caching strategy of data object $k$ at $i$ \\
$L^\text{d}_{k}, L^\text{c}_{mk}$ & size of data $k$ and result size of computation $m$ on $k$  \\
$W_{imk}$ & workload of computation $m$ on data $k$ at node $i$\\
$F_{ij}, D_{ij}(\cdot)$ & total flow rate and transmission cost on link $(i,j)$\\
$G_i, C_i(\cdot)$ & total workload and computation cost at node $i$\\
$Y_i, B_i(\cdot)$ & total cache size and cache deployment cost at $i$\\
$T$ & network aggregated cost\\
$\delta^\text{c}_{ij},\delta^\text{d}_{ij}$ & ``modified marginals'' of $T$ over $\phi^\text{c}_{ij}(m,k)$ and $\phi^\text{d}_{ij}(k)$\\
$\gamma^\text{c}_{i},\gamma^\text{d}_{i}$ & ``modified marginals'' of $T$ over $y^\text{c}_{i}(m,k)$ and $y^\text{d}_{i}(k)$\\
\hline
\end{tabular}
\caption{Major notations}
\vspace{-3\baselineskip}
\label{table:notation}
\end{table}


\section{Offline method: $1/2$ approximation}
\label{sec: submodular}
In this section, we present an offline method for \eqref{Objective_relax}.
The method develops Algorithm $1$ of \cite{zhang2024congestion} to consider multipath forwarding, in-network computation, and reuse of computational results.

Notice that \eqref{flow_conservation_relax} uniquely determines $\boldsymbol{y}$ if $\boldsymbol{\phi}$ is given. 
Let $\mathcal{D}_{\boldsymbol{\phi}}$ be the feasible set (down-closed convex polyhedra) of $\boldsymbol{\phi}$,
\begin{small}
\begin{equation*}
\begin{aligned}
    \mathcal{D}_{\boldsymbol{\phi}} = \Bigg\{ \boldsymbol{\phi} \geq \boldsymbol{0} \, : \, \sum\nolimits_{j \in \{0\}\cup\mathcal{V}} \phi^{\text{c}}_{ij}(m,k) \leq 1, \, \forall (m,k) ;  
    \\\sum\nolimits_{j \in \mathcal{V}} \phi^{\text{d}}_{ij}(k) \begin{cases}
        =0, \, \text{if } i \in \mathcal{S}_k
        \\ \leq 1, \, \text{if } i \not\in \mathcal{S}_k
    \end{cases}  \forall k \Bigg\}.
\end{aligned}
\end{equation*}
\end{small}
We denote by $\boldsymbol{y}(\boldsymbol{\phi})$ the corresponding $\boldsymbol{y}$ for $\boldsymbol{\phi} \in \mathcal{D}_{\boldsymbol{\phi}}$, and rewrite the cost minimization problem \eqref{Objective_relax} as the following \emph{caching-offloading gain maximization} problem,
\begin{equation}
        \max_{\boldsymbol{\phi} \in \mathcal{D}_{\boldsymbol{\phi}}} \quad G(\boldsymbol{\phi}) = T_0 - T(\boldsymbol{y}(\boldsymbol{\phi}),\boldsymbol{\phi}),
    \label{Objective_gain}
\end{equation}
where $G(\boldsymbol{\phi})$ is called the caching-offloading gain, and $T_0$ is a constant upper bound given by (we assume $T_0$ is finite) 
\begin{equation}
    T_0 = \max_{\boldsymbol{\phi} \in\mathcal{D}_{\boldsymbol{\phi}}: \boldsymbol{y}(\boldsymbol{\phi}) = \boldsymbol{0}} T(\boldsymbol{0},\boldsymbol{\phi}),
    \label{T_0}
\end{equation}
namely, the maximum aggregated cost when no data and results are cached. 
We next provide a $1/2$ approximation algorithm to \eqref{Objective_gain} by characterizing and exploiting its mathematical structure.

\subsection{A ``submodular + concave'' reformulation}
\label{subsec: submodular reformulation}
We say there is a \emph{CI path} $p = [p_1,p_2,\cdots,p_{|p|}]$ from node $i$ to node $j$ for computation $m$ on data $k$ if $p_1 = i$, $p_{|p|} = j$, and for any $t = 1,\cdots,|p|-1$, it holds that $(p_t,p_{t+1}) \in \mathcal{E}$ and $\phi^{\text{c}}_{p_t p_{t+1}}(m,k) > 0$.
Similarly, we say $p$ is a \emph{DI path} for data $k$ if $\phi^{\text{d}}_{p_t p_{t+1}}(k) > 0$ for $t = 1,\cdots,|p|-1$.
Let $\mathcal{P}^\text{c}_{ij}(m,k)$ be the set of all CI paths from node $i$ to node $j$ for $m$ and $k$, and let $\mathcal{P}^\text{d}_{ij}(k)$ be the set of all DI paths from $i$ to $j$ for $k$.\footnote{We assume all CI and DI paths are loop-free, i.e., $p_t \neq p_{t^\prime}$ if $t \neq t^\prime$. Mechanisms to guarantee loop-free property are discussed in Section \ref{subsec:GeneralAlgorithm}.}
The rate of CI for $m$ and $k$ generated by node $v$ and arriving at $i$ is $r_v(m,k)\sum_{p \in \mathcal{P}^\text{c}_{vi}(m,k)} \prod_{t = 1}^{|p|-1} \phi^{\text{c}}_{p_t p_{t+1}}(m,k)$, thus 
\begin{subequations}
\label{submod_f_cal}
\begin{small}
\begin{equation}
\begin{aligned}
   & f_{ij}^\text{c}(m,k) = \phi^{\text{c}}_{ij}(m,k)\sum_{v \in \mathcal{V}}r_v(m,k)\sum_{p \in \mathcal{P}^\text{c}_{vi}(m,k)} \prod_{t = 1}^{|p|-1} \phi^{\text{c}}_{p_t p_{t+1}}(m,k),
   \\ & g_i(m,k) = \phi^{\text{c}}_{i0}(m,k)\sum_{v \in \mathcal{V}}r_v(m,k)\sum_{p \in \mathcal{P}^\text{c}_{vi}(m,k)} \prod_{t = 1}^{|p|-1} \phi^{\text{c}}_{p_t p_{t+1}}(m,k).
\end{aligned}
\end{equation}
\end{small}
The rate of DI packets for data $k$ generated by node $v$ and arriving node $i$ is $\left(\sum_{m} g_v(m,k)\right)\sum_{p \in \mathcal{P}^\text{d}_{vi}(k)} \prod_{t = 1}^{|p|-1} \phi^{\text{d}}_{p_t p_{t+1}}(k)$, and
\begin{small}
\begin{equation}
    \begin{aligned}
        &  f_{ij}^\text{d}(k) = \phi^{\text{d}}_{ij}(k) \sum_{v \in \mathcal{V}}\left(\sum_{m \in \mathcal{F}} g_v(m,k)\right)\sum_{p \in \mathcal{P}^\text{d}_{vi}(k)} \prod_{t = 1}^{|p|-1} \phi^{\text{d}}_{p_t p_{t+1}}(k).
    \end{aligned}
\end{equation}
\end{small}
\end{subequations}

We rewrite the gain $G(\boldsymbol{\phi})$ as $G(\boldsymbol{\phi}) = M(\boldsymbol{\phi}) + N(\boldsymbol{\phi})$, where 
\begin{small}
\begin{equation*}
    M(\boldsymbol{\phi}) = T_0 - \sum_{(i,j) \in \mathcal{E}} D_{ij}(F_{ij}) - \sum_{i \in \mathcal{V}}C_{i}(G_{i}), \quad N(\boldsymbol{\phi}) = -\sum_{i \in \mathcal{V}}B_{i}(Y_{i}).
\end{equation*}
\end{small}

By \eqref{submod_f_cal}, $F_{ij}$ and $G_i$ are multilinear in $\boldsymbol{\phi}$ with non-negative coefficients. Combined with convex $D_{ij}(\cdot)$, $F_{ij}(\cdot)$, Lemma \ref{lemma:submodular} holds.

\begin{lem}
\label{lemma:submodular}
    Problem \eqref{Objective_gain} is a ``submodular + concave'' maximization problem. 
    Specifically, $M(\boldsymbol{\phi})$ is non-negative monotonic DR-submodular\footnote{DR-submodular function is a continuous generalization of submodular functions with diminishing return. See, e.g., \cite{bian2017guaranteed}, for more information about DR-submodularity.} in $\boldsymbol{\phi}$, and $N(\boldsymbol{\phi})$ is concave in $\boldsymbol{\phi}$.
\end{lem}

\begin{proof}
    See Appendix \ref{proof:lemma:submodular}.
\end{proof}

\subsection{Algorithm with $1/2$ approximation}
Maximization of ``submodular + concave'' functions was first systematically studied by Mitra et al. \cite{mitra2021submodular}.
Problem \eqref{Objective_gain} falls into one of the categories provided \cite{mitra2021submodular}, to which a \emph{Gradient-Combining Frank-Wolfe} algorithm (we present in Algorithm \ref{alg_GCFW}) guarantees a constant factor approximation of $\frac{1}{2}$.

\begin{theo}[Theorem 3.10 \cite{mitra2021submodular}]
\label{theorem:GCFW_guarantee}
Assume $G(\boldsymbol{\phi})$ is L-smooth, i.e., $\nabla G$ is Lipschitz continuous.
For $N > 1$, let $\boldsymbol{\phi}^{\text{out}}$ be the result of Algorithm \ref{alg_GCFW} and $\boldsymbol{\phi}^*$ be an optimal solution to problem \eqref{Objective_gain}, then
\begin{equation*}
    G(\boldsymbol{\phi}^{\text{out}}) \geq \frac{1 - \varepsilon}{2} M(\boldsymbol{\phi}^*) + N(\boldsymbol{\phi}^*) - \varepsilon \cdot O\left(L |\mathcal{V}|^2|\mathcal{F}||\mathcal{C}|\right).
\end{equation*}
where $L$ is the Lipschitz constant
of $\nabla G$.
\end{theo}

\begin{algorithm}[t]
\SetKwRepeat{DoFor}{do}{for}
\SetKwRepeat{DoDuring}{do}{during}
\SetKwRepeat{DoAt}{do}{at}
\SetKwRepeat{DoWhen}{do}{when}
\SetKwInput{KwInput}{Input}
\KwInput{Integer $N > 1$}
Let $\varepsilon = N^{-\frac{1}{3}}$. Let $n = 0$, and $\boldsymbol{\phi}^{(0)} \in \mathcal{D}_{\boldsymbol{\phi}}$ with $\boldsymbol{y}(\boldsymbol{\phi}^{(0)}) = \boldsymbol{0}$.\\
\DoFor{ $n = 0,1,\cdots, N-1$}
{
Let $\boldsymbol{\psi} = \arg\max_{ \boldsymbol{\phi} \in \mathcal{D}_{\boldsymbol{\phi}} }\, \left\langle \boldsymbol{\phi}, \nabla M(\boldsymbol{\phi}^{(n)}) + 2 \nabla N(\boldsymbol{\phi}^{(n)}) \right\rangle$. \label{line_Linear_Programming} \\
Let $\boldsymbol{\phi}^{(n+1)} = (1 - \varepsilon^2) \boldsymbol{\phi}^{(n)} + \varepsilon^2 \boldsymbol{\psi}$.\\
}
Let $\boldsymbol{\phi}^{\text{out}} = \arg\max_{\boldsymbol{\phi} \in \left\{ \boldsymbol{\phi}^{(0)},\cdots,\boldsymbol{\phi}^{(N)}\right\}}\, G(\boldsymbol{\phi})$ .\\
\caption{Gradient-Combining Frank-Wolfe (GCFW)}
\label{alg_GCFW}
\end{algorithm}

Gradient $\nabla N(\boldsymbol{\phi})$ in Algorithm \ref{alg_GCFW} can be calculated as 
\begin{small}
\begin{equation*}
    \frac{\partial N(\boldsymbol{\phi})}{\partial \phi^\text{c}_{ij}(m,k)} = -L^\text{c}_{mk}B^\prime_{i}(Y_{i}), \quad \frac{\partial N(\boldsymbol{\phi})}{\partial \phi^\text{d}_{ij}(k)} = -L^\text{d}_{k}B^\prime_{i}(Y_{i}).
\end{equation*}
\end{small}
The gradient $\nabla M(\boldsymbol{\phi})$ can be calculated in a centralized server by applying the chain rule combined with \eqref{submod_f_cal}. e.g., for $(i,j) \in \mathcal{E}$,
\begin{small}
\begin{equation}
\begin{aligned}
&\frac{\partial M(\boldsymbol{\phi})}{\partial \phi^\text{c}_{ij}(m,k)} 
=  - \sum_{v \in \mathcal{V}}C^\prime_v(G_v) W_{vmk}\frac{\partial g_v(m,k)}{\partial \phi^\text{c}_{ij}(m,k)} +
\\&-\sum_{(v,u)\in\mathcal{E}}D^\prime_{vu}(F_{vu})\left( L^\text{c}_{mk}\frac{\partial f_{uv}^\text{c}(m,k)}{\partial \phi^\text{c}_{ij}(m,k)} + L^\text{d}_{k}\frac{\partial f_{uv}^\text{d}(k)}{\partial \phi^\text{c}_{ij}(m,k)}\right), 
\end{aligned}
\label{nabla_M}
\end{equation}
\end{small}
where $\frac{\partial g_v(m,k)}{\partial \phi^\text{c}_{ij}(m,k)}$, $\frac{\partial f_{uv}^\text{c}(m,k)}{\partial \phi^\text{c}_{ij}(m,k)}$ and $\frac{\partial f_{uv}^\text{d}(k)}{\partial \phi^\text{c}_{ij}(m,k)}$ can be directly obtained by the closed-form expressions \eqref{submod_f_cal}.
Nevertheless, $\frac{\partial M(\boldsymbol{\phi})}{\partial \phi^\text{c}_{ij}(m,k)}$ can also be calculated from $- \frac{\partial T}{\partial \phi^\text{c}_{ij}(m,k)} - \frac{\partial N(\boldsymbol{\phi})}{\partial \phi^\text{c}_{ij}(m,k)}$, where a distributed recursive calculation of $\frac{\partial T}{\partial \phi^\text{c}_{ij}(m,k)}$ is introduced in Section \ref{subsection: marginals}.

The linear programming in Algorithm \ref{alg_GCFW} can be implemented in a distributed manner:\footnote{Here we only present calculation for $\psi^\text{c}_{ij}(m,k)$. Similar calculation applies to $\psi^\text{d}_{ij}(k)$.} 
for any $m$ and $k$, node $i$ sets $\psi^\text{c}_{ij}(m,k) = 0$ for all $j \in \{0\}\cup\mathcal{V}$ if $\frac{\partial M(\boldsymbol{\phi})}{\partial \phi^{\text{c}}_{ij}(m,k)} + 2\frac{ \partial N(\boldsymbol{\phi})}{\partial \phi^{\text{c}}_{ij}(m,k)} < 0$ for all $j$; otherwise, set $\psi^{\text{c}}_{ij}(m,k) = 1$ for $j = \arg\max_{v \in \{0\}\cup\mathcal{V}} \frac{\partial M(\boldsymbol{\phi})}{\partial \phi^{\text{c}}_{iv}(m,k)} + 2\frac{ \partial N(\boldsymbol{\phi})}{\partial \phi^{\text{c}}_{iv}(m,k)}$.

We remark that although upper bound $T_0$ is defined to be the solution of \eqref{T_0} to guarantee positive $M(\boldsymbol{\phi})$, Algorithm \ref{alg_GCFW} operates identically regardless of $T_0$ as it only involves $\nabla M(\boldsymbol{\phi})$ and $\nabla N(\boldsymbol{\phi})$.
For practical simplicity and effectiveness, the network operator can pick the initial state $\boldsymbol{\phi}^{(0)}$ to be the \emph{shortest-extended-path} scheme with no caching, detailed in Section \ref{sec:simulation}.
To our knowledge, Algorithm \ref{alg_GCFW} provides the first constant factor approximation to the joint communication, caching, and computation placement problem in arbitrary networks with nonlinear costs and computation reuse. 

\section{Online adaptive solution}
\label{sec:general condition}
In practical network scenarios, user request patterns (i.e., user request rates $r_{i}(m,k)$) and network status (e.g., link cost $D_{ij}(\cdot)$ can be affected by the temporary link quality) are both not known prior, and can be time-varying. 
Although Algorithm \ref{alg_GCFW} can be implemented distributed, it is offline, non-adaptive, and requires prior knowledge of request patterns and network status.
The network operator may seek a method with stronger practical feasibility.
To this end, this section presents an online, distributed, and adaptive algorithm to \eqref{Objective_relax}, with minimum prior knowledge requirement. 

We tackle problem \eqref{Objective_relax} with the node-based perspective first used in \cite{gallager1977minimum} and followed by \cite{WiOpt22,zhang2024congestion}.
We first present a KKT necessary optimality condition for \eqref{Objective_relax}, then give a modification to the KKT condition that yields a bounded gap from the global optimum.

\subsection{Closed-form marginals}
\label{subsection: marginals}
We start by giving closed-form partial derivatives of $T(\boldsymbol{y},\boldsymbol{\phi})$. Our analysis makes a non-trivial generalization of \cite{WiOpt22,zhang2024congestion} to cache-enabled computing networks.
For caching strategy $\boldsymbol{y}$, it holds that 
\begin{small}
\begin{equation}
    \frac{\partial T}{\partial y^\text{c}_{i}(m,k)} = L^\text{c}_{mk} B^\prime_{i}(Y_{i}), \quad \frac{\partial T}{\partial y^\text{d}_{i}(k)} = L^\text{d}_{k} B^\prime_{i}(Y_{i}).
    \label{pT_py_cache}
\end{equation}
\end{small}

For CI forwarding strategy $\phi^\text{c}_{ij}(m,k)$ with $j \neq 0$, the marginal cost due to increase of $\phi^\text{c}_{ij}(m,k)$ consists of two parts: 
(1) the marginal cost due to increase of $F_{ji}$ since more CR packets are sent from $j$ to $i$, and 
(2) the marginal cost due to increase of $t^\text{c}_j(m,k)$ since node $j$ needs to handle more CI packets.
Both parts scale with node $i$'s traffic $t_i^\text{c}(m,k)$.
Formally, for all $j \in \mathcal{N}(i)$,
\begin{subequations}
\label{pT_pphi_c}
\begin{small}
\begin{equation}
    \frac{\partial T}{\partial \phi^\text{c}_{ij}(m,k)} = t_i^\text{c}(m,k)\left(L^\text{c}_{mk}D^\prime_{ji}(F_{ji}) + \frac{\partial T}{\partial t_j^{\text{c}}(m,k)}\right).
\end{equation}
\end{small}
Similarly for $j = 0$, marginal cost $\partial T/\partial \phi^{\text{c}}_{i0}(m,k)$ consists of the marginal cost due to workload $G_i$ and the marginal cost due to DI packets generation in $t_i^\text{d}(k)$, i.e.,
\begin{small}
\begin{equation}
    \frac{\partial T}{\partial \phi^\text{c}_{i0}(m,k)} = t_i^\text{c}(m,k)\left(W^\text{c}_{imk}C^\prime_{i}(G_{i}) + \frac{\partial T}{\partial t_i^{\text{d}}(k)}\right).
\end{equation}
\end{small}
\end{subequations}

In \eqref{pT_pphi_c}, term $\partial T/\partial t_i^{\text{c}}(m,k)$ is the marginal cost for $i$ to handle unit rate increment of CI packets for computation $m$ and data $k$. 
It is a weighted sum of marginal costs on out-links and local CPU,
\begin{small}
\begin{equation}
\label{pTpt_c}
\begin{aligned}
    \frac{\partial T}{\partial t_i^{\text{c}}(m,k)} &= \sum_{j \in \mathcal{N}(i)} \phi^\text{c}_{ij}(m,k) 
    \left(L^\text{c}_{mk} D^\prime_{ji}(F_{ji}) + \frac{\partial T}{\partial t^\text{c}_j(m,k)}\right)
    \\ &+ \phi^\text{c}_{i0}(m,k) \left(W_{imk}C^\prime_i(G_i) + \frac{\partial T}{\partial t^\text{d}_i(k)}\right).
\end{aligned}
\end{equation}
\end{small}

A similar reasoning applies to the marginal costs for DI forwarding strategy $\phi^\text{d}_{ij}(k)$. Specifically, similar to \eqref{pT_pphi_c}, it holds that
\begin{small}
\begin{equation}
\label{pT_pphi_d}
\begin{aligned}
    \frac{\partial T}{\partial \phi^\text{d}_{ij}(k)} = t_i^\text{d}(k)\left(L^\text{d}_{k}D^\prime_{ji}(F_{ji}) + \frac{\partial T}{\partial t_j^{\text{d}}(k)}\right), \quad \forall j \in \mathcal{N}(i),
\end{aligned}
\end{equation}
\end{small}
where $\partial T/\partial t_i^{\text{d}}(k)$ is the marginal cost for $i$ to handle unit rate increment of DI packets for data $k$, given by
\begin{small}
\begin{equation}
\label{pTpt_d}
\begin{aligned}
    \frac{\partial T}{\partial t_i^{\text{d}}(k)} = \sum_{j \in \mathcal{N}(i)} \phi^\text{d}_{ij}(k) \left(L^\text{d}_{k} D^\prime_{ji}(F_{ji}) + \frac{\partial T}{\partial t_j^\text{d}(k)}\right).
\end{aligned}
\end{equation}
\end{small}

By \eqref{flow_conservation_relax}, the value of $\partial T/\partial t^\text{c}_{i}(m,k)$ and $\partial T/\partial t^\text{d}_{i}(k)$ are implicitly affected by $\boldsymbol{y}$, e.g., it holds that $\partial T/\partial t^\text{c}_i(m,k) = 0$ if or $y^\text{c}_i(m,k) = 1$. 
Namely, caching computation results locally will immediately set the marginal cost for handling the corresponding CIs to $0$.
Moreover, one could calculate $\partial T/\partial t^\text{c}_i(m,k)$ and $\partial T/\partial t^\text{d}_i(k)$ recursively by \eqref{pTpt_c} and \eqref{pTpt_d}, respectively.
The recursive calculation is guaranteed to terminate in finite steps if no loops are formed\footnote{We refer to loops in either CI paths or DI paths. See definitions in Section \ref{subsec: submodular reformulation}.}.
This can be applied to achieve distributed calculation for $\nabla M(\boldsymbol{\phi})$ in Algorithm \ref{alg_GCFW}.
To carry out the recursive calculation, we introduce a two-stage message broadcasting procedure in Section \ref{subsec:GeneralAlgorithm}.

\subsection{KKT condition and modification}
\label{subsec: conditions}
Based on the closed-form marginals, Lemma \ref{Lemma_KKT} provides a set of KKT necessary conditions (see \cite{bertsekas1997nonlinear} for definition) of problem \eqref{Objective_relax}.
\begin{lem}
    Let $(\boldsymbol{y}, \boldsymbol{\phi})$ be an optimal solution to problem \eqref{Objective_relax}, then for all $i,j \in \mathcal{V}$, $m \in \mathcal{F}$, and $k \in \mathcal{C}$,
\begin{small}

\begin{align}
   & t_i^\text{c}(m,k)\left(L^\text{c}_{mk}D^\prime_{ji}(F_{ji}) + \frac{\partial T}{\partial t_j^{\text{c}}(m,k)}\right) 
    \begin{cases}
        = \lambda^\text{c}_{imk}, \, \text{if } \phi^\text{c}_{ij}(m,k) > 0
        \\ \geq \lambda^\text{c}_{imk}, \, \text{if } \phi^\text{c}_{ij}(m,k) = 0
    \end{cases}  \nonumber
\\ &  t_i^\text{c}(m,k)\left(W^\text{c}_{imk}C^\prime_{i}(G_{i}) + \frac{\partial T}{\partial t_i^{\text{d}}(k)}\right)
    \begin{cases}
        = \lambda^\text{c}_{imk}, \, \text{if } \phi^\text{c}_{i0}(m,k) > 0
        \\ \geq \lambda^\text{c}_{imk}, \, \text{if } \phi^\text{c}_{i0}(m,k) = 0
    \end{cases}  \nonumber
\\  &  t_i^\text{d}(k)\left(L^\text{d}_{k}D^\prime_{ji}(F_{ji}) + \frac{\partial T}{\partial t_j^{\text{d}}(k)}\right) 
    \begin{cases}
        = \lambda^\text{d}_{ik}, \, \text{if } \phi^\text{d}_{ij}(k) > 0
        \\ \geq \lambda^\text{c}_{ik}, \, \text{if } \phi^\text{d}_{ij}(k) = 0
    \end{cases}\label{condition_KKT}
\\ & B^\prime_{i}(Y_{i})  
   \begin{cases}
        = \frac{\lambda^\text{c}_{imk}}{ L^\text{c}_{mk}}, \, \text{if } y^\text{c}_{i}(m,k) > 0
        \\ \geq \frac{\lambda^\text{c}_{imk}}{ L^\text{c}_{mk}}, \, \text{if } y^\text{c}_{i}(m,k) = 0
    \end{cases}\,
B^\prime_{i}(Y_{i})  
   \begin{cases}
        = \frac{\lambda^\text{d}_{ik}}{L^\text{d}_{k}}, \, \text{if } y^\text{d}_{i}(k) > 0
        \\ \geq \frac{\lambda^\text{d}_{ik}}{L^\text{d}_{k}}, \, \text{if } y^\text{d}_{i}(k) = 0
    \end{cases} \nonumber
\end{align} 
\end{small}
where $\lambda^\text{c}_{imk}$ and $\lambda^\text{d}_{ik}$ are given by
\begin{small}
\begin{equation*}
\begin{aligned}
    &\lambda^\text{c}_{imk} = \min\left\{ \frac{\partial T}{\partial y^\text{c}_{i}(m,k)}, \min_{j \in \{0\}\cup\mathcal{V}} \frac{\partial T}{\partial \phi^\text{c}_{ij}(m,k)}\right\},
\\ &\lambda^\text{d}_{ik} = \min\left\{ \frac{\partial T}{\partial y^\text{d}_{i}(k)}, \min_{j \in \mathcal{V}} \frac{\partial T}{\partial \phi^\text{d}_{ij}(k)}\right\}.
\end{aligned}
\end{equation*}
\end{small}
\label{Lemma_KKT}
\end{lem}
\begin{proof}
    See Appendix \ref{proof:Lemma:KKT}.
\end{proof}
The KKT condition \eqref{condition_KKT} is not sufficient for global optimality. A counterexample is provided in \cite{gallager1977minimum}.
Specifically, consider the degenerate case when $t^\text{c}_i(m,k) = t^\text{d}_i(k) = 0$. 
Condition \eqref{condition_KKT} always holds for arbitrary forwarding strategy at $i$, as long as $y^\text{c}_i(m,k) = y^\text{d}_i(k) = 0$.
To this end, we propose a modification to \eqref{condition_KKT} that resolves such degenerate cases.
Notice that in \eqref{condition_KKT}, $t^\text{c}_i(m,k)$ and $t^\text{d}_i(k)$ appear repeatedly in conditions regarding $\boldsymbol{\phi}$ for all $j$.
We divide these conditions by $t^\text{c}_i(m,k)$ or $t^\text{d}_i(k)$, respectively.
Such modification leads to a bounded gap from the global optimum.

\begin{theo}
Let $(\boldsymbol{y}, \boldsymbol{\phi})$ be feasible to \eqref{Objective_relax}, and for all $i \in \mathcal{V}$, $m \in \mathcal{F}$, $k \in \mathcal{C}$, \eqref{condition_mod_1} holds for $j \in \{0\} \cup \mathcal{V}$ and \eqref{condition_mod_2} \eqref{condition_mod_3} hold for $j \in \mathcal{V}$,
\begin{subequations}
\label{condition_mod}
\begin{gather}
    \delta^\text{c}_{ij}(m,k) \begin{cases}
        = \delta^\text{c}_{imk}, \,   \text{if } \phi^\text{c}_{ij}(m,k) > 0
        \\ \geq \delta^\text{c}_{imk}, \,   \text{if } \phi^\text{c}_{ij}(m,k) = 0
    \end{cases} ,\, \forall j \in \{0\} \cup \mathcal{V} \label{condition_mod_1}
\\      \delta^\text{d}_{ij}(k)  \begin{cases}
        = \delta^\text{d}_{ik}, \,   \text{if } \phi^\text{d}_{ij}(k) > 0
        \\ \geq \delta^\text{d}_{ik}, \,   \text{if } \phi^\text{d}_{ij}(k) = 0
    \end{cases} ,\, \forall j \in \mathcal{V} \label{condition_mod_2}
\\ \gamma^\text{c}_i(m,k) \begin{cases}
    = \delta^\text{c}_{imk}, \, \text{if } y^\text{c}_i(m,k) > 0
    \\ \geq \delta^\text{c}_{imk}, \, \text{if } y^\text{c}_i(m,k) = 0
\end{cases}
\gamma^\text{d}_i(k) \begin{cases}
    = \delta^\text{d}_{ik}, \, \text{if } y^\text{d}_i(k) > 0
    \\ \geq \delta^\text{d}_{ik}, \, \text{if } y^\text{d}_i(k) = 0
\end{cases} \label{condition_mod_3}
\end{gather}
\end{subequations}
where $\delta^\text{c}_{ij}(m,k)$, $\delta^\text{d}_{ij}(k)$, $\gamma^\text{c}_i(m,k)$ and $\gamma^\text{d}_i(k)$ are ``modified marginals'', i.e., partial derivatives of $T$ divided by $t^\text{c}_i(m,k)$ and $t^\text{d}_i(k)$, namely,
\footnote{In \eqref{condition_mod_marg_1} \eqref{condition_mod_marg_2}, we assume $\delta^\text{c}_{ij}(m,k) = \delta^\text{d}_{ij}(k) = 0$ if $j \not \in \mathcal{N}(i)$. In \eqref{condition_mod_marg_3}, we assume $\gamma^\text{c}_i(m,k) = \infty$ if $t^\text{c}_{i}(m,k) = 0$, and $\gamma^\text{d}_i(k) = \infty$ if $t^\text{d}_{i}(k) = 0$.}
\begin{subequations}
\label{condition_mod_marg}
\begin{align}
 & \delta^\text{c}_{ij}(m,k) = \begin{cases}
    L^\text{c}_{mk}D^\prime_{ji}(F_{ji}) + \frac{\partial T}{\partial t_j^{\text{c}}(m,k)}, \, \text{if } j \neq 0
    \\ W^\text{c}_{imk}C^\prime_{i}(G_{i}) + \frac{\partial T}{\partial t_i^{\text{d}}(k)},\, \text{if } j = 0
\end{cases}    \label{condition_mod_marg_1}
\\& \delta^\text{d}_{ij}(k) = L^\text{d}_{k}D^\prime_{ji}(F_{ji}) + \frac{\partial T}{\partial t_j^{\text{d}}(k)} \label{condition_mod_marg_2}
\\ & \gamma^\text{c}_i(m,k) = \frac{L^\text{c}_{mk}B^\prime_i(Y_i)}{t^\text{c}_i(m,k)}, \quad 
\gamma^\text{d}_i(k) = \frac{L^\text{d}_{k}B^\prime_i(Y_i)}{t^\text{d}_i(k)}, \label{condition_mod_marg_3}
\end{align}
\end{subequations}
and $\delta^\text{c}_{imk}$, $\delta^\text{d}_{ik}$ are the minimum modified marginals given by
\begin{subequations}
\begin{align}
    \delta^\text{c}_{imk} &= \min\left\{\gamma^\text{c}_i(m,k),\min_{j \in \{0\}\cup\mathcal{V}} \delta^\text{c}_{ij}(m,k)\right\},
\\ \delta^\text{d}_{ik} &= \min\left\{\gamma^\text{d}_i(k),\min_{j \in \mathcal{V}} \delta^\text{d}_{ij}(k)\right\}.
\end{align}
\label{cond_marg_min}
\end{subequations}

Let $(\boldsymbol{y}^\dagger,\boldsymbol{\phi}^\dagger)$ be any feasible solution to \eqref{Objective_relax}. 
Then, it holds that
\begin{gather}
\label{bounded_gap}
    T(\boldsymbol{y}^\dagger, \boldsymbol{\phi}^\dagger) - T(\boldsymbol{y}, \boldsymbol{\phi}) \geq \nonumber
    \\\sum_{i\in\mathcal{V},k \in \mathcal{C}} \delta^\text{d}_{ik} \left(y_{i}^{\text{d}}(k) - y^{\text{d}}_{i}(k)^\dagger\right) \left(t^\text{d}_i(k)^\dagger  - t^\text{d}_i(k) \right) +\label{condition_mod_gap}
    \\ \sum_{i\in\mathcal{V}, m \in \mathcal{F}, k \in \mathcal{C}} \delta^\text{c}_{imk} \left(y_{i}^{\text{c}}(m,k) - y^{\text{c}}_{i}(m,k)^\dagger\right) \left(t^\text{c}_i(m,k)^\dagger  - t^\text{c}_i(m,k) \right). \nonumber
\end{gather}
\label{thm_condition_mod}
\end{theo}
\begin{proof}
    See Appendix \ref{proof:thm_condition_mod}.
\end{proof}
The bounded gap \eqref{bounded_gap} follows the fact that $\sum_i B_i(Y_i)$ is convex in $\boldsymbol{y}$, and $\sum_{i,j}D_{ij}(F_{ij}) + \sum_i C_i(G_i)$ is geodesically convex \footnote{
Geodesic convexity is a generalization of convexity on Riemannian manifolds. See, e.g., \cite{boumal2023introduction}, for the definitions and optimization techniques.} on $\boldsymbol{\phi}$ under additional assumptions.

\subsection{Interpretation and corollaries}
To provide an intuitive interpretation of condition \eqref{condition_mod}, 
recall \eqref{pTpt_c}, the modified marginal for $j \in \mathcal{N}(i)$ in \eqref{condition_mod_marg_1} can be written as 
\begin{equation*}
\begin{aligned}
    \delta_{ij}^\text{c}(m,k) = \frac{\partial T}{\partial \phi^\text{c}_{ij}(m,k)} \bigg/ \frac{\partial f_{ij}^\text{c}(m,k)}{\partial \phi^\text{c}_{ij}(m,k)} = \frac{\partial T}{\partial f_{ij}^\text{c}(m,k)}
\end{aligned}
\end{equation*}
i.e., the marginal cost if node $i$ forwards additional CI packets of unit rate to node $j$.
Similarly, $\delta^\text{c}_{i0}(m,k)$ and $\delta^\text{d}_{ij}(k)$ can be written as
\begin{equation*}
\begin{aligned}
    \delta_{i0}^\text{c}(m,k) = \frac{\partial T}{\partial g_{i}(m,k)}, \quad 
    \delta_{ij}^\text{d}(k) = \frac{\partial T}{\partial f_{ij}^\text{d}(k)}, 
\end{aligned}
\end{equation*}
namely, the marginal cost with respect to computational input at node $i$, and the marginal cost due to additional DI packets of unit rate forwarded to node $j$, respectively. 

On the other hand, we define virtual \emph{cached flows} as $h_{i}^\text{c}(m,k) = t_i^\text{c}(m,k) y_i^\text{c}(m,k)$ and $h_{i}^\text{d}(k) = t_i^\text{d}(k) y_i^\text{d}(k)$, i.e., the rate of CI/DI packets that terminate at node $i$ due to $i$'s caching strategy. Then
\begin{equation}
    \gamma^\text{c}_i(m,k) = \frac{\partial T}{\partial h^\text{c}_{i}(m,k)}, \quad \gamma^\text{d}_i(k) = \frac{\partial T}{\partial h^\text{d}_{i}(k)}
\label{delta_i0_intuitive}
\end{equation}
gives the marginal cache deployment cost for $i$ to resolve unit rate of CI/DI packets by expanding its cache, respectively.

Therefore, \eqref{cond_marg_min} implies that each node independently handles incremental arrival CI/DI packets in the way that achieves the minimum modified marginal -- either by forwarding to neighbors, or by expanding its own cache.
In other words, we say it is ``worthwhile'' to cache result for computation $m$ for data $k$ at node $i$ if $\gamma_i^\text{c}(m,k) < \min_{j \in \{0\} \cup \mathcal{N}(i)} \delta^\text{c}_{ij}(m,k)$, and ``not worth'' otherwise.

Although condition \eqref{condition_mod} is neither a necessary condition nor a sufficient condition for optimality, the set of $(\boldsymbol{y},\boldsymbol{\phi})$ satisfying \eqref{condition_mod} must have a non-empty intersection with the global optima of \eqref{Objective_relax}.
\begin{cor}
For any optimal solution $(\boldsymbol{y}^*,\boldsymbol{\phi}^*)$ to \eqref{Objective_relax}, there exist a $(\boldsymbol{y},\boldsymbol{\phi})$ satisfying \eqref{condition_mod} such that $\boldsymbol{y} = \boldsymbol{y}^*$ and $\phi^\text{c}_{ij}(m,k) = \phi^\text{c}_{ij}(m,k)^*$ for $i,m,k$ with $t_i^\text{c}(m,k)^* > 0$, $\phi^\text{d}_{ij}(k) = \phi^\text{d}_{ij}(k)^*$ for $i,k$ with $t_i^\text{d}(k)^* > 0$.
\label{cor_cache_existance}
\end{cor}


\begin{cor}
Let $(\boldsymbol{y},\boldsymbol{\phi})$ satisfy \eqref{condition_mod}.
Let $(\boldsymbol{y}^\dagger,\boldsymbol{\phi}^\dagger)$ be feasible to \eqref{Objective_relax} such that: for all $i \in \mathcal{V}$, $m \in \mathcal{F}$, $k \in \mathcal{C}$, either $y^\text{c}_{i}(m,k)^\dagger = y^\text{c}_{i}(m,k)$ or $t^\text{c}_{i}(m,k)^\dagger = t^\text{c}_{i}(m,k)$;
for all $i \in \mathcal{V}$ and $k \in \mathcal{C}$, either $y^\text{d}_{i}(k)^\dagger = y^\text{d}_{i}(k)$ or $t^\text{d}_{i}(k)^\dagger = t^\text{d}_{i}(k)$.
Then $T(\boldsymbol{y},\boldsymbol{\phi}) \leq T(\boldsymbol{y}^\dagger,\boldsymbol{\phi}^\dagger)$.
\label{cor_cache_1}
\end{cor}
Corollary \ref{cor_cache_1} implies that \eqref{condition_mod} is sufficient for optimal $\boldsymbol{\phi}$ when $\boldsymbol{y}$ is fixed, and for optimal $\boldsymbol{y}$ when $t_i^\text{c}(m,k)$ and $t^\text{d}_i(k)$ are unchanged.

\begin{figure}[htbp]
\centerline{\includegraphics[width=0.8\linewidth]{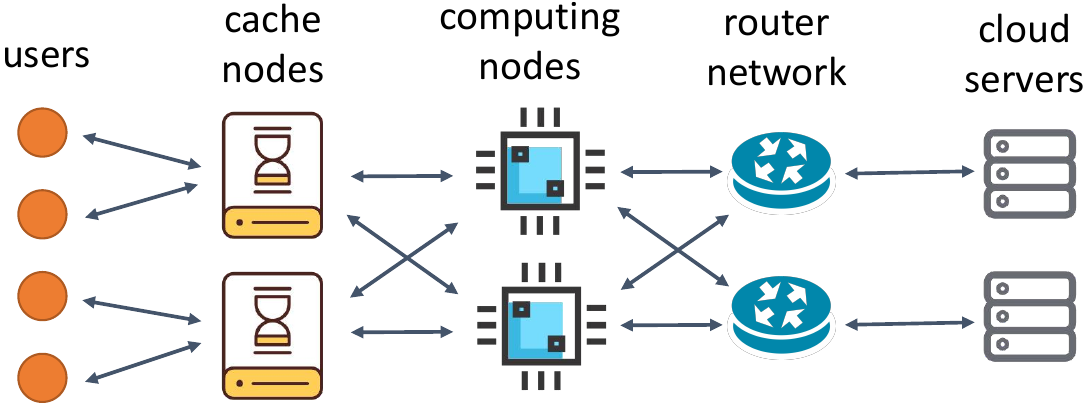}}
\caption{Sample network satisfying Corollary \ref{cor_cache_1}. Single-layered caches and computing nodes are equipped near users. Requests are routed to servers if not fulfilled at the caches.
}
\label{fig_cor2}
\end{figure}

\begin{cor}
Let $(\boldsymbol{y},\boldsymbol{\phi})$ satisfy \eqref{condition_mod}.
Let $(\boldsymbol{y}^\dagger, \boldsymbol{\phi}^\dagger)$ be feasible to \eqref{Objective_relax} such that either $\boldsymbol{\phi}^\dagger \geq \boldsymbol{\phi}$ or $\boldsymbol{\phi}^\dagger \leq \boldsymbol{\phi}$.\footnote{ For two vectors $\boldsymbol{v}_1$ and $\boldsymbol{v}_2$ of the same dimension, we denote by $\boldsymbol{v}_1 \geq \boldsymbol{v}_2$ if every element of $\boldsymbol{v}_1$ is no less than the corresponding element in $\boldsymbol{v}_2$. Similarly as $\boldsymbol{v}_1 \leq \boldsymbol{v}_2$.}
Then $T(\boldsymbol{y},\boldsymbol{\phi}) \leq T(\boldsymbol{y}^\dagger,\boldsymbol{\phi}^\dagger)$.
\label{cor_cache_2}
\end{cor}
If we let $\rho^\text{c}_{ij}(m,k) = \phi^\text{c}_{ij}(m,k) / (1 - y^\text{c}_i(m,k))$ and $\rho^\text{d}_{ij}(k) = \phi^\text{d}_{ij}(k) / (1 - y^\text{d}_i(k))$ be the \emph{conditional forwarding strategy}, i.e., the probability of a CI/DI packet being forwarded to $j$ given that the requested result/data is not cached at $i$.
In practical networks, forwarding and caching mechanisms are usually implemented separately, where the forwarding is only based on the conditional forwarding strategy $\boldsymbol{\rho}$ instead of $\boldsymbol{\phi}$.
Consider the case where $\boldsymbol{\rho}^\dagger = \boldsymbol{\rho}$ but $\boldsymbol{y}^\dagger \geq \boldsymbol{y}$ (or $\boldsymbol{y}^\dagger \leq \boldsymbol{y}$), Corollary \ref{cor_cache_2} implies $T^\dagger \leq T$, i.e., the network aggregated cost cannot be lowered by only caching more items (i.e., only increasing elements in $\boldsymbol{y}$), or only removing items from caches (i.e., only decreasing elements in $\boldsymbol{y}$), while keeping $\boldsymbol{\rho}$ unchanged.

\subsection{Online adaptive algorithm}
\label{subsec:GeneralAlgorithm}
We present a distributed online algorithm that converges to condition \eqref{condition_mod}.
The algorithm does not require prior knowledge of request rates $r_i(m,k)$ and designated servers $\mathcal{S}_k$, and is adaptive to moderate changes in $r_i(m,k)$ and cost functions $D_{ij}(\cdot)$, $B_i(\cdot)$.  $C_i(\cdot)$.

\vspace{0.5\baselineskip}
\noindent\textbf{Algorithm overview.}
The proposed algorithm is a gradient projection variant that generalizes Algorithm 2 in \cite{zhang2024congestion} to cache-enabled computing networks.
We partition time into \emph{slots} of duration $T_{\text{slot}}$, and assume the network starts at a feasible $(\boldsymbol{y}^{(0)}, \boldsymbol{\phi}^{(0)})$ with $T^{(0)} < \infty$.
In $t$-th slot, $\boldsymbol{y}^{(t)}$ and $\boldsymbol{\phi}^{(t)}$ are kept unchanged. At the end of each slot, strategies are updated as
\begin{equation}
\begin{aligned}
 \boldsymbol{\phi}^{(t+1)} = \boldsymbol{\phi}^{(t)} + \Delta\boldsymbol{\phi}^{(t)},
     \quad \boldsymbol{y}^{(t+1)} = \boldsymbol{y}^{(t)} + \Delta \boldsymbol{y}^{(t)},
\end{aligned}
\label{variable_update}
\end{equation}
with distributed calculation for update vectors $\Delta\boldsymbol{\phi}^{(t)}$ and $\Delta\boldsymbol{y}^{(t)}$,
\begin{small}
\begin{equation}
\begin{aligned}
    \Delta\phi_{ij}^{\text{c}}(m,k)^{(t)} &= \begin{cases}
    - \phi_{ij}^{\text{c}}(m,k)^{(t)}, \quad \text{if } j \in \mathcal{B}_i^{\text{c}}(k)^{(t)}
    \\ -\min\left\{ \phi_{ij}^{\text{c}}(m,k)^{(t)}, \alpha e_{ij}^{\text{c}}(m,k)^{(t)} \right\}, \text{if } e_{ij}^{\text{c}}(m,k)^{(t)} > 0
    \\ - S_i^\text{c}(m,k)^{(t)}, \quad
    \text{for one } j $\text{ such that }$ e_{ij}^{\text{c}}(m,k)^{(t)} = 0
    \end{cases}
    \\ \Delta y_{i}^{\text{c}}(m,k)^{(t)} &= \begin{cases}
    -\min\left\{ y_{i}^\text{c}(m,k)^{(t)}, \alpha e_{i,\text{y}}^\text{c}(k)^{(t)} \right\}, \, \text{if } e_{i,\text{y}}^\text{c}(k)^{(t)} > 0
    \\  - S_i^\text{c}(m,k)^{(t)},
    \,  \text{if }e_{i,\text{y}}^\text{c}(k)^{(t)} = 0  \text{, all }  \Delta\phi_{ij}^{\text{c}}(m,k)^{(t)} 
\leq 0
    \end{cases}
\end{aligned}
\label{dphi_and_dy}
\end{equation}
\end{small}
where $\mathcal{B}_i^{\text{c}}(k)^{(t)}$ is the set of \emph{blocked nodes} to prevent routing loops, $\alpha > 0$ is the stepsize, and
\begin{equation*}
\begin{gathered}
e_{ij}^{\text{c}}(m,k)^{(t)} = \delta_{ij}^\text{c}(m,k)^{(t)} - \delta_{imk}^{\text{c} (t)}, \, e_{i,\text{y}}^\text{c}(k)^{(t)} = \gamma_i^\text{c}(m,k)^{(t)} - \delta_{imk}^{\text{c} (t)},
\\ S_i^\text{c}(m,k)^{(t)} = \sum_{j : \Delta\phi_{ij}^{\text{c}}(m,k)^{(t)} \leq 0 }\Delta\phi_{ij}^{\text{c}}(m,k)^{(t)}
+  \Delta y_{i}^{\text{c}}(m,k)^{(t)}.
\end{gathered}
\end{equation*}
The CI strategies, i.e., $\Delta\phi_{ij}^{\text{d}}(k)^{(t)}$ and $\Delta y_{i}^{\text{d}}(k)^{(t)}$, are updated with a same manner and omitted due to space limitation.
The intuitive idea is (i) do not forward any request to blocked nodes, (ii) if a modified marginal ($\delta^\text{c}_{ij}(m,k)$ or $\gamma_i^\text{c}(m,k)$) is not the minimum, shrink the corresponding forwarding/caching fractions, and (iii) add the shrunk fractions to the minimum marginal direction. 
Algorithm \ref{alg_GP} summarizes the proposed online adaptive algorithm.

\begin{algorithm}[t]
\SetKwRepeat{DoFor}{do}{for}
\SetKwRepeat{DoDuring}{do}{during}
\SetKwRepeat{DoAt}{do}{at}
\SetKwRepeat{DoWhen}{do}{when}
\SetKwInput{KwInput}{Input}
\KwInput{Initial loop-free $(\boldsymbol{y}^{(0)},\boldsymbol{\phi}^{(0)})$, stepsize $\alpha$}
\DoAt{ end of $t$-th slot}
{
Update $\partial T / \partial t^\text{c}_{i}(m,k)$ and $\partial T/ \partial t^\text{d}_i(k)$ by Recursive Marginal Calculation.\\
Updates $\boldsymbol{y}^{(t)}$, $\boldsymbol{\phi}^{(t)}$ by \eqref{variable_update} and \eqref{dphi_and_dy}.\\
Round continuous $\boldsymbol{y}^{(t+1)}$ to discrete $\boldsymbol{x}^{(t+1)}$.\\
}
\caption{Gradient Projection (GP)}
\label{alg_GP}
\end{algorithm}

The blocked node set $\mathcal{B}_i^{\text{c}}(k)^{(t)}$ ensures no loops (in both CI paths and DI paths) are formed throughout the algorithm, provided the initial state $(\boldsymbol{y}^{(0)}, \boldsymbol{\phi}^{(0)})$ is loop-free.
The construction of blocked node sets falls into two catalogs, i.e., the \emph{static} sets and the \emph{dynamic} sets.
Static sets are pre-determined and unchanged throughout the algorithm (e.g., in the FIB construction of ICN). Dynamic sets are dynamically calculated as the algorithm proceeds and can adapt to topology changes (e.g., in ad-hoc or self-organized networks).
Discussion and algorithms for both catalogs are provided in \cite{zhang2024congestion}.

After updating \eqref{dphi_and_dy}, the continuous caching strategy $\boldsymbol{y}^{(t+1)}$ is rounded to caching decision $\boldsymbol{x}^{(t+1)}$ for practical implementation.
We adopt the \emph{Distributed Randomized Rounding} method developed by \cite{zhang2024congestion}, guaranteeing that the expected flow rates, computation workloads and cache sizes meet the relaxed value, and the actual cache size $X_i$ at each node is bounded near the expected value $Y_i$.

\vspace{0.5\baselineskip}
\noindent\textbf{Recursive Marginal Calculation.} Recall \eqref{condition_mod_marg}, besides the status of local link/CPU/storage captured by $D^\prime_{ij}(F_{ij})$, $C^\prime_i(G_i)$ and $B^\prime_i(Y_i)$, the calculation in \eqref{dphi_and_dy} requires the knowledge of $\partial T/\partial t^\text{c}_i(m,k)$ and  $\partial T/\partial t^\text{d}_i(k)$, which have a global dependence and can be recursively calculated by \eqref{pTpt_c}\eqref{pTpt_d}.
We utilize the 2-stage message-broadcasting procedure described in \cite{WiOpt22} to carry out the recursive calculation.
The procedure can be used both for \eqref{dphi_and_dy} in the online adaptive Algorithm \ref{alg_GP}, and for \eqref{nabla_M} in the offline Algorithm \ref{alg_GCFW}.
The underlying idea is that: recursive calculation \eqref{pTpt_d} start from $i$ with $i \in \mathcal{S}_k$ or $y_i^\text{d}(k) = 1$ such that $\partial T/\partial t_i^\text{d}(k) = 0$, and a node $j$ calculates $\partial T/\partial t_j^\text{d}(k)$ after obtaining all $\partial T/\partial t_i^\text{d}(k)$ from all downstream neighbors $i : \phi_{ji}^\text{d}(k) > 0$; 
the  recursive calculation \eqref{pTpt_c} starts after all $\partial T/\partial t^\text{d}_i(k)$ are obtained, and starts at nodes with $y^\text{c}_i(m,k) = 1$ or $\phi_{i0}^\text{c}(m,k) = 1$. 
If no loops are formed, this broadcast is guaranteed to traverse throughout the network and terminate within a finite number of steps.

\vspace{0.5\baselineskip}
\noindent\textbf{Convergence and complexity.}
Algorithm \ref{alg_GP} can be online implemented as it does not require prior knowledge of $r_i(m,k)$, and adapts to changes in $r_i(m,k)$ since $F_{ij}$ and $G_i$ can be directly estimated by the packet number on links and CPU. 
It can also adapt network topology change: whenever a link $(i,j)$ is removed from $\mathcal{E}$, node $i$ only needs to add $j$ to the blocked node set; when link $(i,j)$ is added to $\mathcal{E}$, node $i$ removes $j$ from the blocked node set.\footnote{A new node $v$ can randomly initiate $\boldsymbol{\phi}_v$ with \eqref{flow_conservation_relax}, e.g., to the shortest path next-hop.}

\begin{theo}
\label{thm_convergence}
Assume $(\boldsymbol{y}^{(0)},\boldsymbol{\phi}^{(0)})$ is loop-free with $T^0 < \infty$, and $(\boldsymbol{y}^{(t)},\boldsymbol{\phi}^{(t)})$ is updated by Algorithm \ref{alg_GP} with a sufficiently small stepsize $\alpha$. 
Then, the sequence $\left\{(\boldsymbol{y}^{(t)},\boldsymbol{\phi}^{(t)})\right\}_{t = 0}^{\infty}$ converges to a limit point $(\boldsymbol{y}^{*},\boldsymbol{\phi}^{*})$ that satisfies condition \eqref{condition_mod} with loop-free guaranteed.
\end{theo}

Theorem \ref{thm_convergence} is a straightforward extension of \cite{gallager1977minimum} Theorem 5.
The convergence property of Algorithm \ref{alg_GP} can be further improved by adopting second-order quasi-Newton methods, e.g., \cite{xi2008node}.
We denote by $\mathcal{T}$ the set of tasks (i.e., tuple $(d,m,k)$ with $r_d(m,k) > 0$) in the network. The Recursive Marginal Calculation introduces 
$\left(|\mathcal{C}| + |\mathcal{T}|\right)|\mathcal{E}|$ broadcast messages per slot, with on average amount $\left(|\mathcal{C}| + |\mathcal{T}|\right)/T_{\text{slot}}$ per link/second. 
We assume the broadcast messages are sent in an out-of-band channel.
Let $t_c$ be the broadcast message transmission time and $\Bar{h}$ be the maximum length of extended paths, the broadcast completion time is at most $\Bar{h}t_c$.
Algorithm \ref{alg_GP} has space complexity $O\left(|\mathcal{C}| + |\mathcal{T}|\right)$ at each node. 
The update may fail if broadcast completion time exceeds $T$, or if $\left(|\mathcal{C}| + |\mathcal{T}|\right)/T_{\text{slot}}$ exceeds the broadcast channel capacity. 
If so, we can use longer slots, or allow some nodes to update every multiple slots.
If $\mathcal{C}$ or $\mathcal{F}$ is too large or infinite, the network operator should apply the algorithm only to the most popular data/computations, as they contribute to the majority of network traffic/workload. The rest can be handled through other simple methods, e.g. LRU/LFU.

\section{Simulation}
\label{sec:simulation}
We simulate the proposed algorithms and other baseline methods in various network scenarios with a packet-level simulator\footnote{Available at https://github.com/JinkunZhang/Elastic-Caching-Networks.git}.

\label{sec:simulator setting}
\vspace{0.5\baselineskip}
\noindent\textbf{Requests.} 
The requester $i$ is uniformly chosen in all $|\mathcal{V}|$ nodes, the required computation $m$ and data $k$ are chosen in the $\mathcal{F}$, $\mathcal{C}$ with a Zipf-distribution of parameter $1.0$, respectively.
The CI generation rates $r_{d}(m,k)$ for all tasks in $\mathcal{T}$ are uniformly random in $[1.0, 5.0]$.
For each task, $d$ generates CI packets for $m,k$ in a Poisson process of rate $r_{d}(m,k)$.
For each data $ k \in \mathcal{C}$, we assume $|\mathcal{S}_k| = 1$ and choose the designated server uniformly in $\mathcal{V}$.

\begin{figure*}[t]
  \begin{minipage}{.25\linewidth}
    \centering
    \begin{footnotesize}
  \begin{tabularx}{\textwidth}{@{\hspace{1pt}}c@{\hspace{1pt}}|@{\hspace{1pt}}c@{\hspace{3pt}}c@{\hspace{3pt}}c@{\hspace{3pt}}c@{\hspace{3pt}}c@{\hspace{3pt}}c@{\hspace{3pt}}c@{\hspace{3pt}}c}
    \toprule
    Topology & $|\mathcal{V}|$  & $|\mathcal{E}|$ & $|\mathcal{C}|$ & $|\mathcal{F}|$ & $|\mathcal{T}|$ & $\Bar{d}$ & $\Bar{c}$ & $\Bar{b}$  \\
    \midrule
    \texttt{ER} & 50 & 240 & 100 & 20 & 200 & 5 & 10 & 20 \\
    \texttt{grid-100} & 100 & 358 & 100 & 20 & 400 & 5 & 15 & 30  \\
\texttt{Tree} & 63 & 124 & 100 & 20 & 100 & 5 & 10 & 20  \\
\texttt{Fog} & 40 & 130 & 100 & 20 & 150 & 3 & 10 & 30  \\
\texttt{GEANT} & 22 & 66 & 50 & 10 & 100 & 3 & 5 & 10  \\
\texttt{LHC} & 16 & 62 & 50 & 10 & 100  & 3 & 10 & 15 \\
\texttt{DTelekom} & 68 & 546 & 200 & 30 & 400 & 5 & 15 & 20 \\
\texttt{SW} & 120 & 687 & 200 & 30 & 400 & 5 & 15 & 20  \\
    \bottomrule
  \end{tabularx}
\end{footnotesize}
    \captionof{table}
      { Network Scenarios
        \label{tab:scenarios_LOAM}
      }
  \end{minipage}\hfill
  \begin{minipage}{.25\linewidth}
    \centering
    \vspace{-0.5\baselineskip}
    \includegraphics[width=0.98\textwidth]{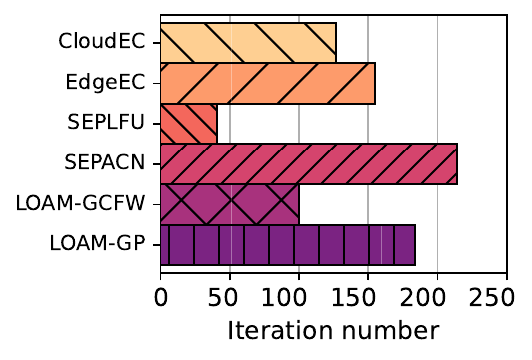}%
    \vspace{-0.6\baselineskip}
    \caption
      {%
        Run time
        \label{fig:convergence}%
      }%
  \end{minipage}\hfill
    \begin{minipage}{.25\linewidth}
    \centering
    \vspace{-0.5\baselineskip}
    \includegraphics[width=0.98\textwidth]{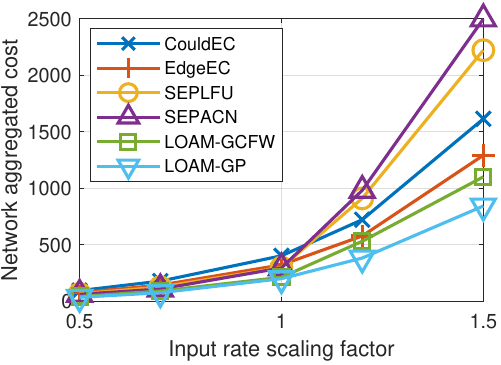}%
    \vspace{-0.7\baselineskip}
    \caption
      {%
        $T$ versus $\alpha$
        \label{fig:trend_inputrate}%
      }%
  \end{minipage}\hfill
  \begin{minipage}{.25\linewidth}
    \centering
    \vspace{-0.5\baselineskip}
    \includegraphics[width=0.98\textwidth]{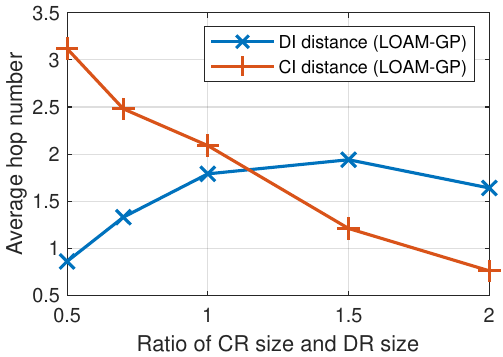}%
    \vspace{-0.7\baselineskip}
    \caption
      {%
        $T$ versus $\beta$
        \label{fig:trend_packetsize}%
      }%
  \end{minipage}
\vspace{-1\baselineskip}
\end{figure*}

\vspace{0.5\baselineskip}
\noindent\textbf{Measurements.}
We monitor the network status at time points with interval $T_{\text{monitor}} = 10$. 
Flows $F_{ij}$ are measured with the average CI/DI packets traveled though $(i,j)$ during past $T_{\text{monitor}}$. 
We use congestion-dependent transmission cost function $D_{ij}(F_{ij}) = F_{ij}/ (1/d_{ij} - F_{ij})$ to capture the average queueing delay on links. 
The computation cost function is also a congestion-dependent average queueing delay at CPUs $C_i(G_i) = G_i/(1/c_i - G_i)$.
Parameter $d_{ij}$ and $c_i$ represent the transmission and computation capacities on links and CPUs.
For methods only considering linear link costs (e.g., when calculating the extended-shortest-path), we use the marginal cost $D_{ij}^\prime(0) = d_{ij}$ and $C_{i}^\prime(0) = c_{i}$ for the weights on links and CPUs, representing linear cost with no congestion.
Cache sizes $Y_i$ are measured as snapshot count of cached items at the monitor time points, and cache deployment cost $B_i(Y_i) = b_i Y_i$, where $b_i$ is the unit cache price at $i$.
Parameters $d_{ij}$, $c_i$, and $b_i$ are uniformly selected from the interval $[0.5 \Bar{d}, 1.5 \Bar{d}]$, $[0.5 \Bar{c}, 1.5 \Bar{c}]$, and $[0.5 \Bar{b}, 1.5 \Bar{b}]$, respectively. The mean value $\Bar{d}$, $\Bar{c}$ and $\Bar{b}$ are given in Table \ref{tab:scenarios_LOAM}.
We assume the data size $L^\text{d}_k = 0.2$ for all $k$, and result size $L^\text{c}_{mk} = 0.1$ for all $m,k$. We assume $W_{imk} = 1$ for all $i$, $m$, $k$. 

\begin{figure*}[h]
  \centering
  \includegraphics[width=0.80\linewidth]{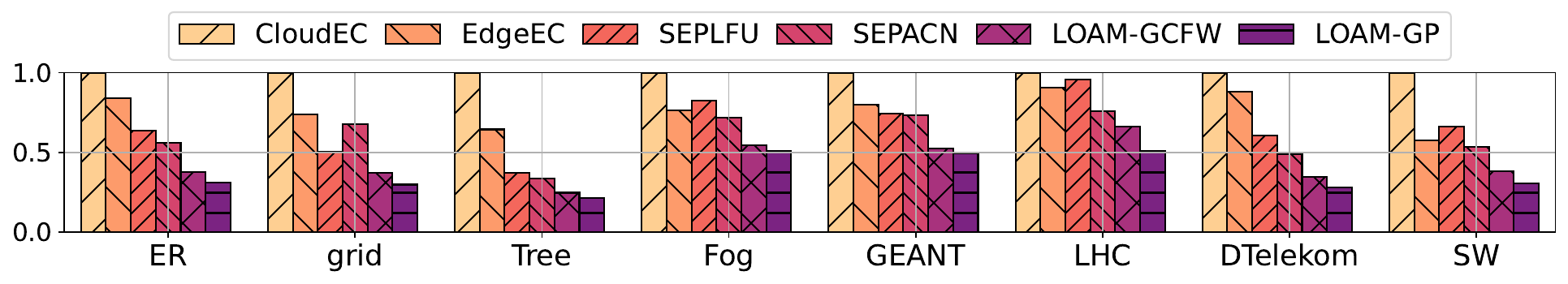}
  \vspace{-0.8\baselineskip}
  \caption{Normalized total cost $T$ of different methods in multiple network scenarios}
 \label{fig_bar_LOAM}
\end{figure*}

\vspace{0.5\baselineskip}
\noindent\textbf{Simulated scenarios and baselines.}
We simulate multiple synthetic or real-world network scenarios summarized in Table \ref{tab:scenarios_LOAM}.
\textbf{\texttt{ER}} is a connectivity-guaranteed Erd\H{o}s-R\'enyi graph, where bi-directional links exist for each pair of nodes with probability $p = 0.07$. 
\textbf{\texttt{grid-100}} and \textbf{\texttt{grid-25}} are $2$-dimensional $10\times10$ and $5\times5$ grid networks. 
\textbf{\texttt{Tree}} is a full binary tree of depth $6$. 
\textbf{\texttt{Fog}} is a full $3$-ary tree of depth $4$, where children of the same parent is concatenated linearly \cite{kamran2021deco}.
\textbf{\texttt{GEANT}} is a pan-European data network for the research and education community \cite{rossi2011caching}.
\textbf{\texttt{LHC}} (Large Hadron Collider) is a prominent data-intensive computing network for high-energy physics applications.
\textbf{\texttt{DTelekom}} is a sample topology of Deutsche Telekom company \cite{rossi2011caching}.
\textbf{\texttt{SW}} (Watts-Strogatz small world) is a ring-graph with additional short-range and long-range edges.

We implement proposed Algorithm \ref{alg_GCFW} (\textbf{\texttt{LOAM-GCFW}}) and Algorithm \ref{alg_GP} (\textbf{\texttt{LOAM-GP}}), and multiple baseline methods: 
\textbf{\texttt{CloudEC}} refers to could computing and elastic caching. Computation requests are sent to the nearest computing server (i.e., nodes with top $5\%$ computation capacity), and Elastic Caching\footnote{Elastic Caching refers to Algorithm 2 in \cite{zhang2024congestion}.} is used to cache computation results.
\textbf{\texttt{EdgeEC}} refers to edge computing and elastic caching. Computations are performed at requester nodes, and Elastic Caching is used to cache data objects.
\textbf{\texttt{SEPLFU}} refers to Shortest Extended Path (SEP) and LFU. SEP is a generalization of the shortest path from requester to data server, where links are weighted by $1/d_{ij}$, and computation is considered a link with weight $1/c_i$. Least Frequently Used (LFU) is used for cache eviction, and cache size is determined by MinCost\footnote{MinCost is a cache deployment algorithm. It adds the cache capacity by $1$ at the node with the highest cache miss cost in every time slot \cite{zhang2024congestion}.}.
\textbf{\texttt{SEPACN}} refers to SEP and Adaptive Caching with Network-wide capacity constraint (ACN)\footnote{ACN is a joint cache deployment and content placement method with a network-wide cache capacity budget \cite{mai2019optimal}.}. We add the total cache budget by $1$ and re-run ACN in each slot.
We set $T_{\text{slot} = 10}$. Both \texttt{LOAM-GCFW} and \texttt{LOAM-GP} start with the shortest extended path and $\boldsymbol{y}^0 = \boldsymbol{0}$.
For \texttt{SEPLFU} and \texttt{SEPACN}, we run simulation for enough long time and record the lowest slot total cost. 
For other methods, we measure steady-state total costs after convergence. 
For \texttt{LOAM-GCFW}, we set $N = 100$. 
For \texttt{LOAM-GP}, we set stepsize $\alpha = 0.01$.

\vspace{0.5\baselineskip}
\noindent\textbf{Results and analysis.}
We summarize the network aggregated costs $T$ across different scenarios in Fig.~\ref{fig_bar_LOAM}. The costs for each scenario are normalized w.r.t. the worst method. 
Fig.~\ref{fig_bar_LOAM} shows that the second group outperforms the first group, and is outperformed by the third group.
This implies that the proposed offline method \texttt{LOAM-GCFW} and online method \texttt{LOAM-GP} outperform other methods in all scenarios, with a total cost reduction of up to $35\%$. We credit such improvement to the fundamental advantages of LOAM's model: the congestion-dependent cost functions, the hop-by-hop routing scheme, and the modeling data/result caching.
Although \texttt{LOAM-GCFW} has a stronger analytical guarantee of $1/2$ approximation, Fig.~\ref{fig_bar_LOAM} shows that \texttt{LOAM-GP} outperforms \texttt{LOAM-GCFW} up to $15\%$.

Fig.~\ref{fig:convergence} compares the convergent iteration number of different methods in \texttt{GEANT}.
The iteration number of \texttt{LOAM-GCFW} is $N$ specified in Algorithm \ref{alg_GCFW}; of \texttt{LOAM-GP}, \texttt{CloudEC}, \texttt{EdgeEC} are measured by the slot number until steady state; of \texttt{SEPLFU}, \texttt{SEPACN} are measured by the slot number until reaching the minimum $T$ (recall that the total cache size is increase by $1$ for each slot).
By setting a desirable $N$, the network operator can straightforwardly adjust the system-established time of \texttt{LOAM-GCFW}, compared to \texttt{LOAM-GP}.

In \texttt{GEANT}, we scale all CI input rates $r_i(m,k)$ by a global factor $\alpha$. 
Fig.~\ref{fig:trend_inputrate} shows the change of total cost $T$ over different $\alpha$, with other parameters fixed. The performance advantage of the proposed methods grows as the network becomes more congested, especially against non-congestion-aware methods, e.g. \texttt{SEPLFU}.

We scale the computation result packet sizes in \texttt{GEANT} and let $\beta$ be the ratio of $L^\text{c}_{mk}$ over $L^\text{d}_k$. In Fig.~\ref{fig:trend_packetsize}, we compare the average travel distance (hop number) of CI and DI packets (i.e., the distance for computation offloading and data retrieval) obtained by \texttt{LOAM-GP} over different $\beta$ with other fixed parameters.
The trajectories suggest that as the result size grows larger, \texttt{LOAM-GP} tends to offload tasks to computation sites at a shorter distance while leaving a longer distance for data retrieval. 
Nevertheless, as the result size grows, the total distance (i.e., the sum of the hop numbers for CI and DI) decreases since a larger result size will generally trigger larger caches and more caching of computation results.

\section{Conclusion}
\label{sec:conclusion}
In this paper, we introduced LOAM, a framework engineered to optimize resource allocation in heterogeneous dispersed computing networks, addressing the limitations of current methods. 
By integrating communication, caching, and computation placement, LOAM offers robust solutions to the NP-hard aggregated cost minimization problem, showcasing superior performance through extensive simulations. 
Our findings reveal LOAM's potential to significantly enhance the efficiency of data- and computation-intensive applications, setting a new benchmark for future research in dispersed computing optimization. 
The adaptability and performance guarantees of LOAM underscore its relevance in evolving network environments, promising to drive advancements in the field.

\bibliographystyle{ACM-Reference-Format}
\bibliography{sample-base}

\appendix

\section{Proof of Lemma \ref{lemma:submodular}}
\label{proof:lemma:submodular}
Recall the definition of $T_0$ and the fact that when one element of $\boldsymbol{\phi}$ is increased while others are unchanged, link flows $F_{ij}$ and computation workloads $G_i$ are non-decreasing for all links and nodes, respectively. Thus the non-negativity and monotonicity of $M(\boldsymbol{\phi})$ are evident. 
By \eqref{flow_conservation_relax}, it holds that
\begin{equation*}
    \begin{aligned}
        y^\text{c}_i(m,k) = 1 - \sum_{j \in \{0\} \cup \mathcal{N}(i)} \phi^{\text{c}}_{ij}(m,k),  \quad     y^\text{d}_i(k) = 1 - \sum_{j \in \mathcal{N}(i)} \phi^{\text{d}}_{ij}(k).
    \end{aligned}
\end{equation*}
Since $B_i(Y_i)$ is convex in $\boldsymbol{y}$, we know $\sum_{i}B_i(Y_i)$ is convex in $\boldsymbol{\phi}$, and the concavity of $N(\boldsymbol{\phi})$ follows.
We prove the DR-submodularity of $M(\boldsymbol{\phi})$ by showing that\footnote{This criteria can be found in \cite{bian2017guaranteed}}
\begin{equation}
    \frac{\partial^2 M(\boldsymbol{\phi})}{\partial \phi_1 \partial \phi_2 } \leq 0,
    \label{fixroute:proof_submod_goal}
\end{equation}
where $\phi_1$ and $\phi_2$ are two elements in vector $\boldsymbol{\phi}$.
For link $(u,v)$ and task $(m,k)$ we have
\begin{equation*}
\begin{aligned}
\frac{\partial M(\boldsymbol{\phi})}{\partial \phi^{\text{c}}_{uv}(m,k)} = - \sum_{(i,j)} \frac{\partial D_{ij}(F_{ij})}{\partial \phi^{\text{c}}_{uv}(m,k)}  - \sum_{i } \frac{\partial C_i(G_i)}{\partial \phi^{\text{c}}_{uv}(m,k)}
\\ = - \sum_{(i,j)} D^\prime_{ij}(F_{ij})\frac{\partial F_{ij}}{\partial \phi^{\text{c}}_{uv}(m,k)}  - \sum_{i } C^\prime_i(G_i) \frac{\partial G_i}{\partial \phi^{\text{c}}_{uv}(m,k)}
\end{aligned}
\end{equation*}
where 
\begin{equation*}
\begin{aligned}
    \frac{\partial F_{ji}}{\partial \phi^{\text{c}}_{uv}(m,k)} = L_{mk}^\text{c}\frac{\partial f_{ij}^\text{c}(m,k)}{\partial \phi^{\text{c}}_{uv}(m,k)} + L_{k}^\text{d}\frac{\partial f_{ij}^\text{d}(k)}{\partial \phi^{\text{c}}_{uv}(m,k)},
\end{aligned}
\end{equation*}
and with link $(u^\prime,v^\prime)$ and task $(m^\prime,k^\prime)$,
\begin{equation*}
\begin{aligned}
    \frac{\partial^2 F_{ji}}{\partial \phi^{\text{c}}_{uv}(m,k) \partial \phi^{\text{c}}_{u^\prime v^\prime}(m^\prime,k^\prime)} = L_{mk}^\text{c}\frac{\partial^2 f_{ij}^\text{c}(m,k)}{\partial \phi^{\text{c}}_{uv}(m,k)\partial \phi^{\text{c}}_{u^\prime v^\prime}(m^\prime,k^\prime)} 
    \\ + L_{k}^\text{d}\frac{\partial^2 f_{ij}^\text{d}(k)}{\partial \phi^{\text{c}}_{uv}(m,k)\partial \phi^{\text{c}}_{u^\prime v^\prime}(m^\prime,k^\prime)}.
\end{aligned}
\end{equation*}
Recall by \eqref{submod_f_cal}, we know $f^\text{c}_{uv}(m,k)$ and $f^\text{d}_{uv}(k)$ are both multilinear in $\boldsymbol{\phi}$ with non-negative coefficients.
Therefore, it holds that ${\partial^2 F_{ji}}/{\partial \phi^{\text{c}}_{uv}(m,k) \partial \phi^{\text{c}}_{u^\prime v^\prime}(m^\prime,k^\prime)}$ is also 
multilinear in $\boldsymbol{\phi}$ with a non-negative coefficient, and so as ${\partial^2 G_{i}}/{\partial \phi^{\text{c}}_{uv}(m,k) \partial \phi^{\text{c}}_{u^\prime v^\prime}(m^\prime,k^\prime)}$.
Note also that by our assumption, 
\begin{equation*}
    D^\prime_{ij}(F_{ij}) \geq 0, \quad D^{\prime\prime}_{ij}(F_{ij}) \geq 0, \quad C^\prime_{i}(G_{i}) \geq 0, \quad C^{\prime\prime}_{i}(G_{i}) \geq 0,
\end{equation*}
Thus it holds that
\begin{equation*}
    \frac{\partial^2 M(\boldsymbol{\phi})}{\partial \phi^{\text{c}}_{uv}(m,k)\partial \phi^{\text{c}}_{u^\prime v^\prime}(m^\prime,k^\prime)} \leq 0.
\end{equation*}
The same reasoning applies for cases when one or both of $\phi_1$, $\phi_2$ are data forwarding strategies $\phi^\text{d}_{uv}(k)$, which completes the proof of DR-submodularity of $M(\boldsymbol{\phi})$.

\section{Proof of Lemma \ref{Lemma_KKT}}
\label{proof:Lemma:KKT}
Note that when \eqref{flow_conservation_relax} holds, $\boldsymbol{\phi} \leq \boldsymbol{1}$ and $\boldsymbol{y} \leq \boldsymbol{1}$ automatically hold. Thus we only consider domain $\boldsymbol{\phi} \geq \boldsymbol{0}$ and $\boldsymbol{y} \geq \boldsymbol{0}$.
The Lagrangian function of problem \ref{Objective_relax} is given by
\begin{equation*}
\begin{aligned}
&L(\boldsymbol{y},\boldsymbol{\phi},\boldsymbol{\lambda},\boldsymbol{\mu}) = T(\boldsymbol{y},\boldsymbol{\phi}) 
\\ &- 
\sum_{i \in \mathcal{V}} \sum_{m, \in \mathcal{F}}\sum_{k \in \mathcal{C}} \lambda^\text{c}_{imk}\left(y^\text{c}_i(m,k) + \sum_{j \in \{0\}\cup\mathcal{V}}\phi^\text{c}_{ij}(m,k) - 1\right)
\\ &- \sum_{i \in \mathcal{V}} \sum_{k \in \mathcal{C}} \lambda^\text{d}_{ik}\left(y^\text{d}_i(k) + \sum_{j \in \mathcal{V}} \phi^\text{d}_{ij}(k) - \mathbbm{1}_{ i \in \mathcal{S}_k}\right)
\\ &- \sum_{i \in \mathcal{V}}\sum_{j \in \mathcal{V}} \sum_{m \in \mathcal{F}}\sum_{k \in \mathcal{C}} \mu^{\text{c,}\phi} \phi^\text{c}_{ij}(m,k) - \sum_{i \in \mathcal{V}}\sum_{m \in \mathcal{F}} \sum_{k \in \mathcal{C}} \mu^{\text{c,}y} y^\text{c}_{i}(m,k)
\\ &- \sum_{i \in \mathcal{V}}\sum_{j \in \mathcal{V}}\sum_{k \in \mathcal{C}} \mu^{\text{d,}\phi} \phi^\text{d}_{ij}(k) - \sum_{i \in \mathcal{V}}\sum_{k \in \mathcal{C}} \mu^{\text{d,}y} y^\text{d}_{i}(k),
\end{aligned}
\end{equation*}
where $\boldsymbol{\lambda} = [\lambda^\text{c}_{imk}, \lambda^\text{d}_{ik}]$ and $\boldsymbol{\mu} = [\mu^{\text{c,}\phi}_{ijmk}, \mu^{\text{c,}y}_{imk}, \mu^{\text{d,}\phi}_{ijk}, \mu^{\text{d,}y}_{ik}]$ are the Lagrangian multipliers for constraints \eqref{flow_conservation_relax} and $\boldsymbol{\phi} \geq \boldsymbol{0}$, $\boldsymbol{y} \geq \boldsymbol{0}$, respectively, and it holds that $\boldsymbol{\mu} \geq \boldsymbol{0}$.
Moreover, by \cite{bertsekas1997nonlinear}, the following complementary slackness holds for all $i$, $j$, $m$ and $k$,
\begin{equation*}
\begin{aligned}
    \mu^{\text{c,}\phi}_{ijmk} \phi^\text{c}_{ij}(m,k) = 0&, \quad \mu^{\text{c,}y}_{imk} y^\text{c}_i(m,k) = 0,
    \\\mu^{\text{d,}\phi}_{ijk} \phi^\text{d}_{ij}(k) = 0&, \quad \mu^{\text{d,}y}_{ik} y^\text{d}_i(k) = 0.
\end{aligned}
\end{equation*}

By letting the derivatives of $L$ with respect to $\boldsymbol{y}$ and $\boldsymbol{\phi}$ equal $0$, we obtain the following KKT conditions,
\begin{equation}
\label{proof:KKT_1}
\begin{aligned}
    \frac{\partial T}{\partial \phi^\text{c}_{ij}(m,k)} = \lambda^\text{c}_{imk} + \mu^{\text{c,}\phi}_{ijmk},& \quad \frac{\partial T}{\partial y^\text{c}_i(m,k)} = \lambda^\text{c}_{imk} + \mu^{\text{c,}y}_{imk},
    \\ \frac{\partial T}{\partial \phi^\text{d}_{ij}(k)} = \lambda^\text{d}_{ik} + \mu^{\text{d,}\phi}_{ijk},& \quad \frac{\partial T}{\partial y^\text{d}_i(k)} = \lambda^\text{d}_{ik} + \mu^{\text{d,}y}_{ik}.
\end{aligned} 
\end{equation}
In other words, the existence of multipliers $\boldsymbol{\lambda}$, $\boldsymbol{\mu}$ so that \eqref{proof:KKT_1} holds is a necessary condition for $(\boldsymbol{y},\boldsymbol{\phi})$ to optimally solve $\eqref{Objective_relax}$. 
Note that each element of multiplier $\boldsymbol{\mu}$ appears only once in \eqref{proof:KKT_1}, then the existence of such $\boldsymbol{\lambda}$, $\boldsymbol{\mu}$ reduces to the existence of $\boldsymbol{\lambda}$ such that
\begin{equation}
\label{proof:KKT_2}
\begin{aligned}
    \frac{\partial T}{\partial \phi^\text{c}_{ij}(m,k)} 
    \begin{cases}
        = \lambda^\text{c}_{imk}, \, \text{if } \phi^\text{c}_{ij}(m,k) > 0
        \\ \geq \lambda^\text{c}_{imk}, \, \text{if } \phi^\text{c}_{ij}(m,k) = 0
    \end{cases},    
\\ \frac{\partial T}{\partial y^\text{c}_i(m,k)}  
   \begin{cases}
        = \lambda^\text{c}_{imk}, \, \text{if } y^\text{c}_{i}(m,k) > 0
        \\ \geq \lambda^\text{c}_{imk}, \, \text{if } y^\text{c}_{i}(m,k) = 0
    \end{cases},
\\    \frac{\partial T}{\partial \phi^\text{d}_{ij}(k)} 
    \begin{cases}
        = \lambda^\text{d}_{ik}, \, \text{if } \phi^\text{d}_{ij}(k) > 0
        \\ \geq \lambda^\text{c}_{ik}, \, \text{if } \phi^\text{d}_{ij}(k) = 0
    \end{cases},    
\\ \frac{\partial T}{\partial y^\text{d}_i(k)}  
   \begin{cases}
        = \lambda^\text{d}_{ik}, \, \text{if } y^\text{d}_{i}(k) > 0
        \\ \geq \lambda^\text{d}_{ik}, \, \text{if } y^\text{d}_{i}(k) = 0
    \end{cases}.
\end{aligned} 
\end{equation}
Substituting the closed-form partial derivatives \eqref{pT_pphi_c} \eqref{pT_pphi_d} into \eqref{proof:KKT_2}, and notice the arbitrariness of $\boldsymbol{\lambda}$, \eqref{proof:KKT_2} is equivalent to the following,
\begin{equation*}
\begin{aligned}
   & t_i^\text{c}(m,k)\left(L^\text{c}_{mk}D^\prime_{ji}(F_{ji}) + \frac{\partial T}{\partial t_j^{\text{c}}(m,k)}\right) 
    \begin{cases}
        = \lambda^\text{c}_{imk}, \, \text{if } \phi^\text{c}_{ij}(m,k) > 0
        \\ \geq \lambda^\text{c}_{imk}, \, \text{if } \phi^\text{c}_{ij}(m,k) = 0
    \end{cases}  
\\ &  t_i^\text{c}(m,k)\left(W^\text{c}_{imk}C^\prime_{i}(G_{i}) + \frac{\partial T}{\partial t_i^{\text{d}}(k)}\right)
    \begin{cases}
        = \lambda^\text{c}_{imk}, \, \text{if } \phi^\text{c}_{i0}(m,k) > 0
        \\ \geq \lambda^\text{c}_{imk}, \, \text{if } \phi^\text{c}_{i0}(m,k) = 0
    \end{cases}  
\\  &  t_i^\text{d}(k)\left(L^\text{d}_{k}D^\prime_{ji}(F_{ji}) + \frac{\partial T}{\partial t_j^{\text{d}}(k)}\right) 
    \begin{cases}
        = \lambda^\text{d}_{ik}, \, \text{if } \phi^\text{d}_{ij}(k) > 0
        \\ \geq \lambda^\text{c}_{ik}, \, \text{if } \phi^\text{d}_{ij}(k) = 0
    \end{cases}
\\ & B^\prime_{i}(Y_{i})  
   \begin{cases}
        = \frac{\lambda^\text{c}_{imk}}{ L^\text{c}_{mk}}, \, \text{if } y^\text{c}_{i}(m,k) > 0
        \\ \geq \frac{\lambda^\text{c}_{imk}}{ L^\text{c}_{mk}}, \, \text{if } y^\text{c}_{i}(m,k) = 0
    \end{cases}\,
B^\prime_{i}(Y_{i})  
   \begin{cases}
        = \frac{\lambda^\text{d}_{ik}}{L^\text{d}_{k}}, \, \text{if } y^\text{d}_{i}(k) > 0
        \\ \geq \frac{\lambda^\text{d}_{ik}}{L^\text{d}_{k}}, \, \text{if } y^\text{d}_{i}(k) = 0
    \end{cases}
\end{aligned} 
\end{equation*}
where $\lambda^\text{c}_{imk}$ and $\lambda^\text{d}_{ik}$ are given by
\begin{equation*}
\begin{aligned}
    &\lambda^\text{c}_{imk} = \min\left\{ \frac{\partial T}{\partial y^\text{c}_{i}(m,k)}, \min_{j \in \{0\}\cup\mathcal{V}} \frac{\partial T}{\partial \phi^\text{c}_{ij}(m,k)}\right\},
\\ &\lambda^\text{d}_{ik} = \min\left\{ \frac{\partial T}{\partial y^\text{d}_{i}(k)}, \min_{j \in \mathcal{V}} \frac{\partial T}{\partial \phi^\text{d}_{ij}(k)}\right\}.
\end{aligned}
\end{equation*}

\section{Proof of Theorem \ref{thm_condition_mod}}
\label{proof:thm_condition_mod}
Our proof combines and generalizes the methods by \cite{WiOpt22} and \cite{zhang2024congestion}.
We start with the flows for data transmission.
Since when $i \in \mathcal{S}_k$, constraint \eqref{flow_conservation_relax} implies $\phi^\text{d}_{ij}(k)$ and $y^\text{d}_i(k)$ are all $0$ which automatically satisfies \eqref{condition_mod}, we only consider the case with $y^\text{d}_i(k) + \sum_{j}\phi^\text{d}_{ij}(k) = 1$.
Note that when condition \eqref{condition_mod} holds, for link $(i,j)$ with $\phi^\text{d}_{ij}(k) > 0$,
\begin{equation}
    L^\text{d}_{k}D^\prime_{ji}(F_{ji}) + \frac{\partial T}{\partial t^\text{d}_j(k)} = \delta^\text{d}_{ik}.
    \label{Proof_cache_1}
\end{equation}
Multiply \eqref{Proof_cache_1} by $\phi^\text{d}_{ij}(k)$ and sum over all such $j$, we have
\begin{equation*}
    \sum_{j: \phi^\text{d}_{ij}(k) > 0} \phi^\text{d}_{ij}(k)\left(L^\text{d}_k D^\prime_{ji}(F_{ji}) + \frac{\partial T}{\partial t^\text{d}_j(k)}\right) = \delta^\text{d}_{ik} \sum_{j: \phi_{ij}(k) > 0}\phi^\text{d}_{ij}(k).
\end{equation*}
Define $p^\text{d}_{ik} = \sum_{j\in\mathcal{N}(j)}\phi^\text{d}_{ij}(k)$ and combining with \eqref{pT_pphi_d}, then
\begin{equation}
    \frac{\partial T}{\partial t^\text{d}_i(k)} = p^\text{d}_{ik}\delta^\text{d}_{ik}.
\label{Proof_cache_1.5}
\end{equation}
By condition \eqref{condition_mod}, for all $j \in \mathcal{N}(i)$,
\begin{equation}
    L^\text{d}_k D^\prime_{ji}(F_{ji}) + \frac{\partial T}{\partial t^\text{d}_j(k)} \geq \delta^\text{d}_{ik}.
\label{Proof_cache_2}
\end{equation}
To bring in the arbitrarily chosen $(\boldsymbol{y}^\dagger,\boldsymbol{\phi}^\dagger)$, multiply \eqref{Proof_cache_2} by $\phi^\text{d}_{ij}(k)^\dagger$ and sum over all $j \in \mathcal{N}(i)$, we have
\begin{equation}
    \sum_{j \in \mathcal{N}(i)} \left(L^\text{d}_k D^\prime_{ji}(F_{ji}) + \frac{\partial T}{\partial t^\text{d}_j(k)}\right)\phi^\text{d}_{ij}(k)^\dagger \geq \delta^\text{d}_{ik}\left(\sum_{j \in \mathcal{N}(i)}\phi^\text{d}_{ij}(k)^\dagger\right).
\label{Proof_cache_2.5}
\end{equation}
Let $p^{\text{d}\dagger}_{ik} = \sum_{j \in \mathcal{N}(i)}\phi^\text{d}_{ij}(k)^\dagger$ and rearrange the terms, we have
\begin{equation}
    \sum_{j \in \mathcal{N}(i)} L^\text{d}_kD^\prime_{ji}(F_{ji})\phi^\text{d}_{ij}(k)^\dagger
    \geq p^{\text{d}\dagger}_{ik} \delta^\text{d}_{ik} - \sum_{j \in \mathcal{N}(i)}\phi^\text{d}_{ij}(k)^\dagger \frac{\partial T}{\partial t^\text{d}_j(k)}.
\label{Proof_cache_3}
\end{equation}
Multiply \eqref{Proof_cache_3} by $t^\text{d}_{i}(k)^\dagger$, let $f^\text{d}_{ij}(k)^\dagger = t^\text{d}_{i}(k)^\dagger\phi^\text{d}_{ij}(k)^\dagger$, we have
\begin{equation}
\begin{aligned}
    &\sum_{j \in \mathcal{N}(i)} L^\text{d}_k D^\prime_{ji}(F_{ji})f_{ij}^\text{d}(k)^\dagger 
    \\& \geq p^{\text{d}\dagger}_{ik} t^\text{d}_{i}(k)^\dagger \delta^\text{d}_{ik}
    -  \sum_{j \in \mathcal{N}(i)}\phi^\text{d}_{ij}(k)^\dagger t^\text{d}_i(k)^\dagger \frac{\partial T}{\partial t^\text{d}_j(k)}.
\end{aligned}
\label{Proof_cache_4}
\end{equation}

Let $F_{ij}^{\text{d}} = \sum_{k}L^\text{d}_k f^\text{d}_{ji}(k)$ and sum \eqref{Proof_cache_4} over $i$ and $k$, we have
\begin{equation}
\begin{aligned}
&\sum_{i \in \mathcal{V}}\sum_{j\in\mathcal{N}(i)} D^\prime_{ji}(F_{ji}) F_{ji}^{\text{d}\dagger} \geq \sum_{i\in\mathcal{V}}\sum_{k \in \mathcal{C}}p_{ik}^{\text{d}\dagger} t^\text{d}_i(k)^\dagger \delta^\text{d}_{ik} 
\\&- \sum_{i \in \mathcal{V}}\sum_{j \in \mathcal{N}(i)}\sum_{k \in \mathcal{C}}\phi^\text{d}_{ij}(k)^\dagger t^\text{d}_i(k)^\dagger \frac{\partial T}{\partial t^\text{d}_j(k)}.
\label{Proof_cache_5}
\end{aligned}
\end{equation}

Recall that $\sum_{i \in \mathcal{N}(j)}\phi^\text{d}_{ij}(k)^\dagger t^\text{d}_i(k)^\dagger = t^\text{d}_j(k)^\dagger - \sum_{m}g_j(m,k)^\dagger$, we swap the summation order in the last term of \eqref{Proof_cache_5}, and replace it by
\begin{equation}
\begin{aligned}
    &- \sum_{j \in \mathcal{V}}\sum_{k \in \mathcal{C}}\frac{\partial T}{\partial t^\text{d}_j(k)}\left(\sum_{i \in \mathcal{N}(j)}\phi^\text{d}_{ij}(k)^\dagger t^\text{d}_i(k)^\dagger\right)
    \\&= \sum_{i \in \mathcal{V}}\sum_{k \in \mathcal{C}}\frac{\partial T}{\partial t^\text{d}_i(k)}\left(\sum_{m\in\mathcal{F}}g_i(m,k)^\dagger - t^\text{d}_i(k)^\dagger\right).
\end{aligned}
\label{Proof_cache_5.1}
\end{equation}

Therefore, \eqref{Proof_cache_5} is equivalent to
\begin{equation}
\begin{aligned}
    &\sum_{i \in \mathcal{V}}\sum_{j\in\mathcal{N}(i)} D^\prime_{ji}(F_{ji}) F^{\text{d}\dagger}_{ji} \geq
    \sum_{i\in\mathcal{V}}\sum_{k \in \mathcal{C}}p_{ik}^{\text{d}\dagger} t^\text{d}_i(k)^\dagger \delta^\text{d}_{ik} 
    \\&+ \sum_{i \in \mathcal{V}}\sum_{k \in \mathcal{C}}\frac{\partial T}{\partial t^\text{d}_i(k)}\left(\sum_{m\in\mathcal{F}}g_i(m,k)^\dagger - t^\text{d}_i(k)^\dagger\right).
\end{aligned}
\label{Proof_cache_5.2}
\end{equation}

Recall \eqref{Proof_cache_1.5} and replace $\partial T/ \partial t^\text{d}_i(k)$, the above is equivalent to
\begin{equation}
\begin{aligned}
    &\sum_{i \in \mathcal{V}}\sum_{j\in\mathcal{N}(i)} D^\prime_{ji}(F_{ji})  F^{\text{d}\dagger}_{ji} \geq
    \sum_{i\in\mathcal{V}}\sum_{k \in \mathcal{C}} \left(p_{ik}^{\text{d}\dagger} -p_{ik}^{\text{d}}\right)t^\text{d}_i(k)^\dagger\delta^\text{d}_{ik} 
    \\& + \sum_{i \in \mathcal{V}}\sum_{k \in \mathcal{C}}\frac{\partial T}{\partial t^\text{d}_i(k)}\left(\sum_{m \in \mathcal{F}}g_i(m,k)^\dagger\right).
\end{aligned}
\label{Proof_cache_6}
\end{equation}


To derive an analog of \eqref{Proof_cache_6} for $\boldsymbol{\phi}^\dagger = \boldsymbol{\phi}$, \eqref{pTpt_d} is equivalent to
\begin{equation*}
    \sum_{j \in \mathcal{N}(i)} L^\text{d}_k D^\prime_{ji}(F_{ji}) \phi^\text{d}_{ij}(k)= \frac{\partial T}{\partial t^\text{d}_i(k)} - \sum_{j \in \mathcal{N}(i)}\phi^\text{d}_{ij}(k)\frac{\partial T}{\partial t^\text{d}_j(k)}.
\end{equation*}
Multiply the above by $t^\text{d}_i(k)$ and sum over $i$ and $k$, we have
\begin{equation*}
\begin{aligned}
    &\sum_{i \in \mathcal{V}} \sum_{j \in \mathcal{N}(i)} D^\prime_{ji}(F_{ji}) F^\text{d}_{ji} = \sum_{i \in \mathcal{V}} \sum_{k \in \mathcal{C}}t^\text{d}_i(k) \frac{\partial T}{\partial t^\text{d}_i(k)} 
    \\&- \sum_{i \in \mathcal{V}} \sum_{k \in \mathcal{C}} \sum_{j \in \mathcal{N}(i)} \phi^\text{d}_{ij}(k)t^\text{d}_i(k)\frac{\partial T}{\partial t^\text{d}_j(k)}.
\end{aligned}
\end{equation*}
Replace the last term with $\sum_{i}\sum_{k} \left(\sum_{m}g_i(m,k) - t^\text{d}_i(k)\right)\partial T / \partial t^\text{d}_i(k)$,
\begin{equation}
    \sum_{i \in \mathcal{V}} \sum_{j \in \mathcal{N}(i)} D^\prime_{ji}(F_{ji}) F^\text{d}_{ji} =
     \sum_{i \in \mathcal{V}} \sum_{k \in \mathcal{C}} \frac{\partial T}{\partial t^\text{d}_i(k)}\left(\sum_{m \in \mathcal{F}}g_i(m,k)\right).
\label{Proof_cache_7}
\end{equation}

Subtract \eqref{Proof_cache_7} from \eqref{Proof_cache_6}, we have 
\begin{equation}
\begin{aligned}
&\sum_{(j,i) \in \mathcal{E}} D^\prime_{ji}(F_{ji}) \left(F_{ji}^{\text{d}\dagger} - F^\text{d}_{ji}\right)
\geq  
\sum_{i\in\mathcal{V}}\sum_{k \in \mathcal{C}} \left(p_{ik}^{\text{d}\dagger} - p^\text{d}_{ik}\right)t^\text{d}_i(k)^\dagger \delta^\text{d}_{ik}
\\ &+ \sum_{i \in \mathcal{V}}\sum_{k \in \mathcal{C}}\frac{\partial T}{\partial t^\text{d}_i(k)}\left(\sum_{m \in \mathcal{F}}g_i(m,k)^\dagger - \sum_{m \in \mathcal{F}}g_i(m,k)\right).
\end{aligned}
\label{Proof_cache_8}
\end{equation}

Consider caching for data contents. By condition \eqref{condition_mod}, we have
\begin{equation*}
     {L^\text{d}_{k}B^\prime_i(Y_i)} \geq t^\text{d}_i(k)\delta^\text{d}_{ik}, 
\end{equation*}
and the equality holds when $y_{i}(k) > 0$.
Let $Y^\text{d}_i = \sum_{k}L^\text{d}_k y^\text{d}_i(k)$, then 
\begin{equation}
\begin{aligned}
        \sum_{i\in\mathcal{V}} B^\prime_{i}(Y_{i})\left(Y_{i}^{\text{d}\dagger} - Y^\text{d}_{i}\right) 
    &=  \sum_{i\in\mathcal{V}} \sum_{k \in \mathcal{C}} B^\prime_{i}(Y_{i})L^\text{d}_k\left(y_{i}^{\text{d}\dagger}(k) - y^\text{d}_{i}(k)\right) 
    \\& \geq \sum_{i \in \mathcal{V}} \sum_{k \in \mathcal{C}}t^\text{d}_i(k)\delta^\text{d}_{ik} \left(y_{i}^{\text{d}\dagger}(k) - y^\text{d}_{i}(k)\right).
\end{aligned}
\label{Proof_cache_9}
\end{equation}
Note that in the RHS of \eqref{Proof_cache_9}, for the case $i \not \in \mathcal{S}_k$, we must have $y_{i}^{\text{d}\dagger}(k) = 1- p_{ik}^{\text{d}\dagger}$ and $y_{i}^\text{d}(k) = 1- p^\text{d}_{ik}$. For the case $i \in \mathcal{S}_k$, we have $p_{ik}^{\text{d}\dagger} = p^\text{d}_{ik} \equiv 0$ as well as $y_{i}^{\text{d}\dagger}(k) = y^\text{d}_{i}(k) \equiv 0$.
Then \eqref{Proof_cache_9} implies
\begin{equation}
\begin{aligned}
        \sum_{i\in\mathcal{V}} B^\prime_{i}(Y_{i})\left(Y_{i}^{\text{d}\dagger} - Y^\text{d}_{i}\right) 
    \geq \sum_{i\in\mathcal{V}} \sum_{k \in \mathcal{C}}t^\text{d}_i(k)\delta^\text{d}_{ik}\left(p^\text{d}_{ik} - p_{ik}^{\text{d}\dagger}\right).
\end{aligned}
\label{Proof_cache_9.5}
\end{equation}

Combining \eqref{Proof_cache_8} and \eqref{Proof_cache_9.5}, it holds that
\begin{equation}
\begin{aligned}
   & \sum_{(j,i) \in \mathcal{E}} D^\prime_{ji}(F_{ji}) \left(F_{ji}^{\text{d}\dagger} - F^\text{d}_{ji}\right) + \sum_{i\in\mathcal{V}} B^\prime_{i}(Y_{i})\left(Y_{i}^{\text{d}\dagger} - Y^\text{d}_{i}\right) 
   \\& \geq \sum_{i\in\mathcal{V}}\sum_{k \in \mathcal{C}} \delta^\text{d}_{ik} \left(p_{ik}^{\text{d}\dagger} - p^\text{d}_{ik}\right) \left(t^\text{d}_i(k)^\dagger  - t^\text{d}_i(k) \right)
\\ &+ \sum_{i \in \mathcal{V}}\sum_{k \in \mathcal{C}}\frac{\partial T}{\partial t^\text{d}_i(k)}\left(\sum_{m \in \mathcal{F}}g_i(m,k)^\dagger - \sum_{m \in \mathcal{F}}g_i(m,k)\right).
\end{aligned}
\label{Proof_cache_data}
\end{equation}

We apply a similar procedure to caching and forwarding strategies for CI packets.  
Specifically, \eqref{Proof_cache_1} becomes
\begin{equation*}
    L^\text{c}_{mk}D^\prime_{ji}(F_{ji}) + \frac{\partial T}{\partial t^\text{c}_j(m,k)} = \delta^\text{c}_{imk},
\end{equation*}
and if $\phi^\text{c}_{i0}(m,k) > 0$, it holds that
\begin{equation*}
    W_{imk} C^\prime_i(G_i) + \frac{\partial T}{\partial t^\text{d}_i(k)} = \delta^\text{c}_{imk}. 
\end{equation*}
Let $p^\text{c}_{imk} = \sum_{ j \in \{0\}\cup\mathcal{V}} \phi^\text{c}_{ij}(m,k)$, and it holds that
\begin{equation}
    \frac{\partial T}{\partial t^\text{c}_i(m,k)} = p^\text{c}_{imk} \delta^\text{c}_{imk}.
\end{equation}
Then \eqref{Proof_cache_2} becomes
\begin{equation}
\begin{aligned}
    &L^\text{c}_{mk} D^\prime_{ji}(F_{ji}) + \frac{\partial T}{\partial t^\text{c}_{i}(m,k)} \geq \delta^\text{c}_{imk},
    \\&W_{imk} C^\prime_i(G_i) + \frac{\partial T}{\partial t^\text{d}_{i}(k)} \geq \delta^\text{c}_{imk},
\end{aligned}
\end{equation}
and \eqref{Proof_cache_2.5} becomes
\begin{equation}
\begin{aligned}
    &\sum_{j \in \mathcal{N}(i)} \left(L^\text{c}_{mk} D^\prime_{ji}(F_{ji}) + \frac{\partial T}{\partial t^\text{c}_j(m,k)}\right)\phi^\text{c}_{ij}(m,k)^\dagger
    \\& + \left(W_{imk} C^\prime_i(G_i) + \frac{\partial T}{\partial t^\text{d}_i(k)}\right)\phi^\text{c}_{i0}(m,k)^\dagger
    \\& \geq \delta^\text{c}_{imk}\left(\sum_{j \in \{0\} \cup \mathcal{N}(i)}\phi^\text{c}_{ij}(m,k)^\dagger\right).
\end{aligned}
\end{equation}
Then \eqref{Proof_cache_3} becomes
\begin{equation}
\begin{aligned}
    &W_{imk}C^\prime_i(G_i) \phi^\text{c}_{i0}(m,k)^\dagger + \sum_{j \in \mathcal{N}(i)} L^\text{c}_{mk} D^\prime_{ji}(F_{ji})\phi^\text{c}_{ij}(m,k)^\dagger 
    \\& \geq p_{imk}^{\text{c}\dagger}\delta^\text{c}_{imk}
    - \phi^\text{c}_{i0}(m,k)^\dagger \frac{\partial T}{\partial t^\text{c}_i(k)}
    - \sum_{j \in \mathcal{N}(i)}\phi^\text{c}_{ij}(m,k)^\dagger \frac{\partial T}{\partial t^\text{c}_j(m,k)},
\end{aligned}
\end{equation}
and \eqref{Proof_cache_4} becomes
\begin{equation}
\begin{aligned}
    &W_{imk}C^\prime_i(G_i) g_{i}(m,k)^\dagger + \sum_{j \in \mathcal{N}(i)} L^\text{c}_{mk} D^\prime_{ji}(F_{ji})f^\text{c}_{ij}(m,k)^\dagger 
\\& \geq p_{imk}^{\text{c}\dagger}t^\text{c}_i(m,k)^\dagger\delta^\text{c}_{imk} 
- \phi^\text{c}_{i0}(m,k)^\dagger t^\text{c}_i(m,k)^\dagger \frac{\partial T}{\partial t^\text{d}_i(k)}
\\& - \sum_{j \in \mathcal{N}} \phi^\text{c}_{ij}(m,k)^\dagger t^\text{c}_i(m,k)^\dagger \frac{\partial T}{\partial t^\text{c}_j(m,k)},
\end{aligned}
\end{equation}
thus \eqref{Proof_cache_5} becomes
\begin{equation}
\begin{aligned}
    & \sum_{i \in \mathcal{V}} C^\prime_i(G_i)G_i^\dagger + \sum_{i \in \mathcal{V}}\sum_{j \in \mathcal{N}(i)} D^\prime_{ji}(F_{ji}) F^{\text{c}\dagger}_{ji}
    \\& \geq \sum_{i \in \mathcal{V}} \sum_{m \in \mathcal{F}} \sum_{k \in \mathcal{C}} p_{imk}^{\text{c}\dagger}t^\text{c}_i(m,k)^\dagger\delta^\text{c}_{imk} 
    \\&- \sum_{i \in \mathcal{V}} \sum_{m \in \mathcal{F}} \sum_{k \in \mathcal{C}} \phi^\text{c}_{i0}(m,k)^\dagger t^\text{c}_i(m,k)^\dagger \frac{\partial T}{\partial t^\text{d}_i(k)}
    \\& - \sum_{i \in \mathcal{V}}\sum_{j \in \mathcal{N}(i)} \sum_{m \in \mathcal{F}}\sum_{k \in \mathcal{C}} \phi^\text{c}_{ij}(m,k)^\dagger t^\text{c}_i(m,k)^\dagger \frac{\partial T}{\partial t^\text{c}_i(m,k)}.
\end{aligned}
\end{equation}

Moreover, \eqref{Proof_cache_5.1} becomes
\begin{equation}
\begin{aligned}
   & - \sum_{j \in \mathcal{V}} \sum_{m \in \mathcal{F}}\sum_{k \in \mathcal{C}}  \frac{\partial T}{\partial t^\text{c}_j(m,k)} \left( \sum_{i \in \mathcal{N}(j)} \phi^\text{c}_{ij}(m,k)^\dagger t^\text{c}_i(m,k)^\dagger\right)
   \\& = \sum_{i \in \mathcal{V}} \sum_{m \in \mathcal{F}}\sum_{k \in \mathcal{C}}  \frac{\partial T}{\partial t^\text{c}_i(m,k)}\left(r_i(m,k) - t^\text{c}_i(m,k)^\dagger\right),
\end{aligned}
\end{equation}
and \eqref{Proof_cache_5.2} becomes
\begin{equation}
\begin{aligned}
      & \sum_{i \in \mathcal{V}} C^\prime_i(G_i)G_i^\dagger + \sum_{i \in \mathcal{V}}\sum_{j \in \mathcal{N}(i)} D^\prime_{ji}(F_{ji}) F^{\text{c}\dagger}_{ji}
    \\& \geq \sum_{i \in \mathcal{V}} \sum_{m \in \mathcal{F}} \sum_{k \in \mathcal{C}} p_{imk}^{\text{c}\dagger}t^\text{c}_i(m,k)^\dagger\delta^\text{c}_{imk} 
    \\&- \sum_{i \in \mathcal{V}} \sum_{m \in \mathcal{F}} \sum_{k \in \mathcal{C}} \phi^\text{c}_{i0}(m,k)^\dagger t^\text{c}_i(m,k)^\dagger \frac{\partial T}{\partial t^\text{d}_i(k)}
    \\& + \sum_{i \in \mathcal{V}} \sum_{m \in \mathcal{F}}\sum_{k \in \mathcal{C}}  \frac{\partial T}{\partial t^\text{c}_i(m,k)}\left(r_i(m,k) - t^\text{c}_i(m,k)^\dagger\right).
\end{aligned}
\end{equation}

Substitute $\partial T/\partial t^\text{c}_i(m,k)$ and cancel, \eqref{Proof_cache_6} becomes
\begin{equation}
\begin{aligned}
      & \sum_{i \in \mathcal{V}} C^\prime_i(G_i)G_i^\dagger + \sum_{i \in \mathcal{V}}\sum_{j \in \mathcal{N}(i)} D^\prime_{ji}(F_{ji}) F^{\text{c}\dagger}_{ji}
    \\& \geq \sum_{i \in \mathcal{V}} \sum_{m \in \mathcal{F}} \sum_{k \in \mathcal{C}} \left(p^{\text{c}\dagger}_{imk} - p^\text{c}_{imk}\right)\delta^\text{c}_{imk} t^\text{c}_{i}(m,k)^\dagger
    \\& + \sum_{i \in \mathcal{V}} \sum_{m \in \mathcal{F}} \sum_{k \in \mathcal{C}} p^\text{c}_{imk} \delta^\text{c}_{imk} r_{i}(m,k)
    \\& - \sum_{i \in \mathcal{V}} \sum_{m \in \mathcal{F}} \sum_{k \in \mathcal{C}} g_i(m,k)^\dagger \frac{\partial T}{\partial t^\text{d}_i(k)},
\end{aligned}    
\end{equation}
and \eqref{Proof_cache_7} becomes
\begin{equation}
\begin{aligned}
     & \sum_{i \in \mathcal{V}} C^\prime_i(G_i)G_i + \sum_{i \in \mathcal{V}}\sum_{j \in \mathcal{N}(i)} D^\prime_{ji}(F_{ji}) F^{\text{c}}_{ji}
     \\& = \sum_{i \in \mathcal{V}} \sum_{m \in \mathcal{F}} \sum_{k \in \mathcal{C}} p^\text{c}_{imk} \delta^\text{c}_{imk} r_{i}(m,k)
    \\& - \sum_{i \in \mathcal{V}} \sum_{m \in \mathcal{F}} \sum_{k \in \mathcal{C}} g_i(m,k) \frac{\partial T}{\partial t^\text{d}_i(k)}.
\end{aligned}
\end{equation}

Therefore, \eqref{Proof_cache_8} becomes
\begin{equation}
\begin{aligned}
     & \sum_{i \in \mathcal{V}} C^\prime_i(G_i)\left(G_i^\dagger - G_i\right) + \sum_{i \in \mathcal{V}}\sum_{j \in \mathcal{N}(i)} D^\prime_{ji}(F_{ji}) \left(F^{\text{c}\dagger}_{ji} - F^{\text{c}}_{ji}\right)
     \\& \geq \sum_{i \in \mathcal{V}} \sum_{m \in \mathcal{F}} \sum_{k \in \mathcal{C}} \left(p^{\text{c}\dagger}_{imk} - p^\text{c}_{imk}\right)\delta^\text{c}_{imk} t^\text{c}_{i}(m,k)^\dagger
     \\& - \sum_{i \in \mathcal{V}} \sum_{m \in \mathcal{F}} \sum_{k \in \mathcal{C}}\frac{\partial T}{\partial t^\text{d}_i(k)} \left(g_i(m,k)^\dagger - g_i(m,k)\right) 
\end{aligned}
\end{equation}

On the other hand, for caching strategies of computation results, \eqref{Proof_cache_9.5} becomes
\begin{equation}
\begin{aligned}
    \sum_{i \in \mathcal{V}} B^\prime_i(Y_i) \left(Y^{\text{c}\dagger}_i - Y^\text{c}_i\right) \geq \sum_{i\in\mathcal{V}}\sum_{m \in\mathcal{F}}\sum_{k \in \mathcal{C}}t^\text{c}_i(m,k)\delta^\text{c}_{imk}\left(p^{\text{c}}_{imk} - p^{\text{c}\dagger}_{imk}\right),
\end{aligned}
\end{equation}
and therefore \eqref{Proof_cache_data} becomes
\begin{equation}
\begin{aligned}
       & \sum_{(j,i) \in \mathcal{E}} D^\prime_{ji}(F_{ji}) \left(F_{ji}^{\text{c}\dagger} - F^\text{c}_{ji}\right) 
       + \sum_{i \in \mathcal{V}} C^\prime_i(G_i)\left(G_i^\dagger - G_i\right) 
       \\&+ \sum_{i\in\mathcal{V}} B^\prime_{i}(Y_{i})\left(Y_{i}^{\text{c}\dagger} - Y^\text{c}_{i}\right) 
   \\& \geq \sum_{i\in\mathcal{V}}\sum_{m \in \mathcal{F}}\sum_{k \in \mathcal{C}} \delta^\text{c}_{imk} \left(p_{imk}^{\text{c}\dagger} - p^\text{c}_{imk}\right) \left(t^\text{c}_i(m,k)^\dagger  - t^\text{c}_i(m,k) \right)
\\ &- \sum_{i \in \mathcal{V}}\sum_{k \in \mathcal{C}}\frac{\partial T}{\partial t^\text{d}_i(k)}\left(\sum_{m \in \mathcal{F}}g_i(m,k)^\dagger - \sum_{m \in \mathcal{F}}g_i(m,k)\right).
\end{aligned}
\label{Proof_cache_result}
\end{equation}

Summing \eqref{Proof_cache_data} and \eqref{Proof_cache_result}, we have
\begin{equation}
\begin{aligned}
    & \sum_{(j,i) \in \mathcal{E}} D^\prime_{ji}(F_{ji}) \left(F_{ji}^{\dagger} - F_{ji}\right) 
       + \sum_{i \in \mathcal{V}} C^\prime_i(G_i)\left(G_i^\dagger - G_i\right) 
       \\&+ \sum_{i\in\mathcal{V}} B^\prime_{i}(Y_{i})\left(Y_{i}^{\text{c}\dagger} - Y^\text{c}_{i}\right) 
    \\& \geq \sum_{i\in\mathcal{V}}\sum_{k \in \mathcal{C}} \delta^\text{d}_{ik} \left(p_{ik}^{\text{d}\dagger} - p^\text{d}_{ik}\right) \left(t^\text{d}_i(k)^\dagger  - t^\text{d}_i(k) \right)
    \\& + \sum_{i\in\mathcal{V}}\sum_{m \in \mathcal{F}}\sum_{k \in \mathcal{C}} \delta^\text{c}_{imk} \left(p_{imk}^{\text{c}\dagger} - p^\text{c}_{imk}\right) \left(t^\text{c}_i(m,k)^\dagger  - t^\text{c}_i(m,k) \right).
\end{aligned}
\label{Proof_cache_sum}
\end{equation}

Finally we compare $T = (\boldsymbol{y},\boldsymbol{\phi})$ and $T^\dagger = (\boldsymbol{y}^\dagger,\boldsymbol{\phi}^\dagger)$.
Note that $T$ as a function is jointly convex in the total link rates $\boldsymbol{F} = [F_{ij}]_{(i,j)\in\mathcal{E}}$, computation workloads $\boldsymbol{G} = [G_i]_{i \in \mathcal{V}}$, and the occupied cache sizes $\boldsymbol{Y} = [Y_{i}]_{i\in\mathcal{V}}$, due to the convexity of $D_{ij}(\cdot)$, $C_i(\cdot)$, and $B_{m}(\cdot)$.
Thus
\begin{equation}
\begin{aligned}
    T^\dagger - T & \geq \left(\boldsymbol{F}^\dagger - \boldsymbol{F} \right) \nabla_{\boldsymbol{F}}T 
    + \left(\boldsymbol{G}^\dagger - \boldsymbol{G} \right) \nabla_{\boldsymbol{G}}T 
    +\left(\boldsymbol{Y}^\dagger - \boldsymbol{Y} \right) \nabla_{\boldsymbol{Y}}T
    \\ &= \sum_{(i,j) \in \mathcal{E}}\left(F_{ij}^\dagger - F_{ij}\right)D^\prime_{ij}(F_{ij})
    + \sum_{i \in \mathcal{V}} \left(Y_{i}^\dagger - Y_{i}\right)B^\prime_{i}(Y_{i})
    \\ & + \sum_{i \in \mathcal{V}} C^\prime_i(G_i)\left(G_i^\dagger - G_i\right).
\end{aligned}
\label{Proof_cache_10}
\end{equation}

Thus by \eqref{Proof_cache_sum}, we have
\begin{equation*}
\begin{aligned}
    &T^\dagger - T \geq \sum_{i\in\mathcal{V}}\sum_{k \in \mathcal{C}} \delta^\text{d}_{ik} \left(p_{ik}^{\text{d}\dagger} - p^\text{d}_{ik}\right) \left(t^\text{d}_i(k)^\dagger  - t^\text{d}_i(k) \right)
    \\& + \sum_{i\in\mathcal{V}}\sum_{m \in \mathcal{F}}\sum_{k \in \mathcal{C}} \delta^\text{c}_{imk} \left(p_{imk}^{\text{c}\dagger} - p^\text{c}_{imk}\right) \left(t^\text{c}_i(m,k)^\dagger  - t^\text{c}_i(m,k) \right).
\end{aligned}
\end{equation*}
which completes the proof.

We remark that the proof above reflects that $\sum_{i,j}D_{ij}(F_{ij}) + \sum_i C_i(G_i)$ is geodesically convex in $\boldsymbol{\phi}$ when it holds $t_i^\text{c}(m,k) > 0$ for all $i,m,k$ and $t_i^\text{d}(k) > 0$ for all $i,k$.
In such cases, there exists a one-to-one mapping from $\boldsymbol{\phi}$ to $\boldsymbol{f}$, where $\boldsymbol{f} = [f_{ij}^\text{c}(m,k), f_{ij}^\text{d}(k), g_i(m,k)]$ is the flow-domain variable vector.
Moreover, the Jacobian matrix of this mapping is full-rank as the mapping functions are continuously differentiable both from $\boldsymbol{\phi}$ to $\boldsymbol{f}$ and from $\boldsymbol{f}$ to $\boldsymbol{\phi}$.
When the geodesic convexity holds, problem \eqref{Objective_relax} is, in fact, a minimization of the sum of a convex function and a geodesically convex function. To the best of our knowledge, our previous work \cite{zhang2024congestion} is the first to tackle such a problem with an analytical guarantee. The above proof adapts and generalizes the method by \cite{zhang2024congestion} to LOAM network settings.

\section{Proof of Corollary \ref{cor_cache_existance}}
\label{proof_cor_existance}

To prove the existence of a global optimal solution that satisfies condition \eqref{condition_mod}, without loss of generality, we assume the global optimal solution to \eqref{Objective_relax} exist, and with finite objective value $T^* < \infty$. 
In this proof, we show that for any given $(\boldsymbol{y}^*,\boldsymbol{\phi}^*)$ that optimally solves \eqref{Objective_relax} with finite objective value $T^*$, there always exists a $(\boldsymbol{y},\boldsymbol{\phi})$ that satisfies \eqref{condition_mod} and with $T(\boldsymbol{y},\boldsymbol{\phi}) = T^*$. 
To show that, we construct such $(\boldsymbol{y},\boldsymbol{\phi})$ from $(\boldsymbol{y}^*,\boldsymbol{\phi}^*)$.

We first observe that the condition for caching, i.e., \eqref{condition_mod_3}, must be satisfied by $(\boldsymbol{y}^*, \boldsymbol{\phi}^*)$.
In fact, \eqref{condition_mod_3} holds trivially from the KKT condition \ref{condition_KKT} applied to $(\boldsymbol{y}^*,\boldsymbol{\phi}^*)$: for any $i,m,k$, if $t_i^\text{c}(m,k)^* >0$, then \eqref{condition_mod_3} holds as it is equivalent to \eqref{condition_KKT}; 
if $t_i^\text{c}(m,k)^*  = 0$, then we must have $y_{i}^\text{c}(m,k)^* = 0$, as caching computation result for $m,k$ at $i$ will not reduce link and computation costs, but only increase the caching cost; a similar situation happens for cases of $t_i^\text{d}(k)^*  > 0$ and $t_i^\text{d}(k)^*  = 0$.
Hence, we construct the caching variable as 
\begin{equation*}
    \boldsymbol{y} = \boldsymbol{y}^*.
\end{equation*}
We next construct $\boldsymbol{\phi}$. 
Let set 
\begin{equation*}
\begin{aligned}
    &\mathcal{C}^{\text{c}*} = \left\{(i,m,k) \big| t_i^\text{c}(m,k)^* > 0\right\}, \quad \mathcal{C}^{\text{d}*} = \left\{(i,k) \big| t_i^\text{d}(k)^* > 0\right\}, 
\end{aligned}
\end{equation*}
For elements of $\boldsymbol{\phi}$ corresponding to $\mathcal{C}^{\text{c}*}$ and $\mathcal{C}^{\text{d}*}$, we let
\begin{equation}
\begin{aligned}
    &\phi_{ij}^\text{c}(m,k) = \phi_{ij}^\text{c}(m,k)^*, \quad \forall (i,m,k) \in \mathcal{C}^{\text{c}*}, \, \forall j \in \{0\} \cup \mathcal{N}(i),
    \\ &\phi_{ij}^\text{d}(k) = \phi_{ij}^\text{d}(k)^*, \quad \forall (i,k) \in \mathcal{C}^{\text{d}*}, \, \forall j \in \mathcal{N}(i).
\end{aligned}
\label{cor_exi_proof_1}
\end{equation}
For each $(i,m,k) \not\in \mathcal{C}^{\text{c}*}$, we pick one $j \in \{0\} \cup \mathcal{N}(i)$ with
\begin{equation}
    j \in \arg\min_{j^\prime \in \{0\}\cup\mathcal{N}(i)} \delta_{ij}^\text{c}(m,k)
\end{equation}
and let $\phi_{ij}^\text{c}(m,k) = 1$, while let $\phi_{ij^\prime}^\text{c}(m,k) = 0$ for all other $j^\prime \neq j$.
For each $(i,k) \not\in \mathcal{C}^{\text{d}*}$, if $i \in \mathcal{S}_k$, we let $\phi_{ij}^\text{d}(k) = 0$. Otherwise, we pick one $j \in\mathcal{N}(i)$ with
\begin{equation}
    j \in \arg\min_{j^\prime \in \mathcal{N}(i)} \delta_{ij}^\text{d}(k)
\end{equation}
and let $\phi_{ij}^\text{d}(k) = 1$, while let $\phi_{ij^\prime}^\text{d}(k) = 0$ for all other $j^\prime \neq j$.

Such construction is always feasible. For example, the construction could start at sinks (nodes with $y_i^\text{c}(m,k) = 1$ or $y_i^\text{d}(k) = 1$) and destinations (nodes $ i \in \mathcal{S}_k$) and propagates in the upstream order. It is easy to verify that the constructed $(\boldsymbol{y},\boldsymbol{\phi})$ is a global optimal solution to \eqref{Objective_relax} and satisfies \eqref{condition_mod}, which completes the proof.

\section{Proof of Corollary \ref{cor_cache_2}}
\label{proof_cor_f_increase}
We first prove the case $\boldsymbol{\phi}^\dagger \geq \boldsymbol{\phi}$. 
Let $$p_{imk}^\text{c} = 1 - y_{i}^\text{c}(m,k) = \sum_{j \in \{0\} \cup \mathcal{V}} \phi_{ij}^\text{c}(m,k).$$
In this case, it is obvious that 
\begin{equation*}
    p_{imk}^{\text{c}\dagger} \geq p_{imk}^\text{c}, \quad \forall i \in \mathcal{V}, m\in\mathcal{F}, k \in \mathcal{C}.
\end{equation*}
Let $$p_{ik}^\text{d} = 1 - y_{i}^\text{d}(k) = \sum_{j \in \mathcal{V}} \phi_{ij}^\text{d}(k).$$
Then it holds that
\begin{equation*}
    p_{ik}^{\text{d}\dagger} \geq p_{ik}^\text{d}, \quad \forall i \in \mathcal{V}, k \in \mathcal{C}.
\end{equation*}
Meanwhile, since $\phi_{ij}^\text{c}(m,k)^\dagger \geq \phi_{ij}^\text{c}(m,k)$, given the exogenous input rates $r_i(m,k)$ unchanged, the corresponding workloads and link flows for computation results are also non-decreasing. Namely, 
\begin{equation*}
\begin{aligned}
    g_i(m,k)^\dagger &\geq g_i(m,k), \quad \forall i, m, k
\\    f_{ij}^\text{c}(m,k)^\dagger &\geq f_{ij}^\text{c}(m,k), \quad \forall i,j,m,k,
    \\t_i^\text{c}(m,k)^\dagger &\geq t_i^\text{c}(m,k), \quad \forall i,m,k.
\end{aligned}
\end{equation*}
Further, since $\phi_{ij}^\text{d}(k)^\dagger \geq \phi_{ij}^\text{d}(k)$, it holds that
\begin{equation*}
\begin{aligned}
    f_{ij}^\text{d}(k)^\dagger &\geq f_{ij}^\text{d}(k), \quad \forall i,j,k,
    \\t_i^\text{d}(k)^\dagger &\geq t_i^\text{d}(k), \quad \forall i,k.
\end{aligned}
\end{equation*}
Combing the above with Theorem \ref{thm_condition_mod} and noticing $\delta^\text{c}_{imk}$ and $\delta^\text{d}_{ik}$ are both non-negative, we have $T(\boldsymbol{y}^\dagger, \boldsymbol{\phi}^\dagger) \geq T(\boldsymbol{y},\boldsymbol{\phi})$. 
The same reasoning applies for case $\boldsymbol{\phi}^\dagger \leq \boldsymbol{\phi}$, which completes the proof.

\end{document}